\documentclass[a4paper,11pt]{article}
\usepackage{comment}
\usepackage{hyphenat}
\usepackage{array}
\usepackage{listings}
\usepackage{url}
\usepackage{enumitem}

\AtBeginDocument{%
}

\usepackage{graphicx}
\usepackage{tcolorbox} 
\definecolor{mygray}{HTML}{bfbfbf}
\newlength{\topboxpadding}
\newlength{\bottomboxpadding}
\newlength{\leftboxpadding}
\newlength{\rightboxpadding}
\usepackage{subcaption}
\usepackage{authblk}
\usepackage{fancyvrb}
\usepackage{caption} 
\usepackage{hyperref} 
\usepackage{authblk}
\usepackage{comment}
\usepackage[T1]{fontenc}



\usepackage{geometry}

\usepackage{tabularx}
\usepackage{adjustbox}

\newcommand{\Description}[1]{}

\begin{document}

\title{Design of a visual environment for programming by direct data manipulation}

\author[2]{Michel Adam}
\author[1, 3]{Patrice Frison}
\author[3]{Moncef Daoud}
\author[3]{Sabine Letellier Zarshenas}

\affil[1]{IRISA}
\affil[2]{Doctor of Computer Science}
\affil[3]{University of South Brittany}

\date{} 

\maketitle

\begin{abstract}
The use of applications on computers, smartphones, and tablets has been considerably simplified thanks to interactive and dynamic graphical interfaces coupled with the mouse and touch screens. It is no longer necessary to be a computer specialist to use them. Paradoxically, the development of computer programs generally requires writing lines of code in a programming language whose syntax is particularly strict. This process poses many difficulties for programmers. We propose an original tool in which arbitrary programs (Turing-complete) can be developed in a completely visual manner by direct manipulation of the data,
without writing a line of code.
The user can thus develop an algorithm by directly visualizing the result of actions taken on the data.
A method for constructing iterations is associated with the tool.
It proposes to create each part, including the loop body, in a non-linear manner under visual control of the state of the data.
In addition, the tool supports the production of code that corresponds to the actions performed, where the language can be Python, C, or Java.
In this article, we present the tool, the design choices, the problems solved, and the limits and  contributions of the direct-data-manipulation approach.
\end{abstract}

\section{Introduction}
The use of computers, tablets and smartphones has been facilitated by graphical user interfaces. The screen has become the visual support for user actions. Users use their fingers, touchpad, mouse, and touchscreen to manipulate objects displayed on the screen. They can thus move them, make them
disappear or hide them, make new ones appear, transform them, trigger an action/reaction by acting
on these objects, etc. Direct, visual, and largely intuitive manipulation is today at the heart of
human-computer interactions (HCI).

Here we will focus on the paradigm of computer programming through direct visual manipulation of
classic programming objects (variables, constants, arrays, indexes, arithmetic operations,
conditionals, loops, I/O) to:

\begin{enumerate}
\item Develop and implement an algorithm that solves a given problem, \
\item Record the actions and operations performed on or with these objects,
\item Automatically generate the sequence of instructions corresponding to the program thus constructed. \
\end{enumerate}
Our approach is to allow the user, especially a learner, to manipulate programming objects via
a graphical interface, to directly visualize the result of these manipulations, to see the correspondence between the
manipulations carried out and the lines of code produced, and to obtain a directly usable program. Our work
departs from visual programming systems such as Scratch and Blockly \cite{resnick2009scratch}, \cite{trower2015blockly}, \cite{fraser2015ten}
that rely on the use of graphical blocks to assemble instructions and build a program.
Our work is also distinct from the No Code proposal \cite{frank2021low} that employs visual tools to simplify application development for people with no programming skills.
On the contrary, it concerns the design of algorithms and their implementation in code. 

Let us start with a quick trip back in time, when computer scientists typed command lines to perform all the processing.
Graphical interfaces did not yet exist or were just beginning to be
developed.  
The user did not have direct and graphical manipulation tools to automatically translate their actions into command lines. 
Before the use of modern devices, the user had to issue commands themselves on a "terminal". 
For example, to replace the word "search" with "research" at
line 25 of the file "myFile.txt" in the "Doc" folder of the user's personal space, they would type the following sequence:
\begin{center}
\begin{lstlisting}[float= h]
        cd home
        cd Doc
        edit myFile.txt
        25s/search/research/
        w
        q
\end{lstlisting}
\end{center}
This command sequence performs the following operations:
\begin{enumerate}
\item Position in the user’s personal space,
\item Move to the correct folder, \
\item Launch a text editor on the correct file, \
\item Specify the substitution to be made,
\item Save to current file,
\item Exit the text editor.
\end{enumerate}
With a graphical screen, users display their home folder, then the Doc folder. 
They open the file “myFile.txt” to launch the text editor. 
They select the word “search” in the displayed text, which they replace directly with “research”. 
They save their work and close the editor. All of this is done by direct manipulation. 
In fact, the graphical interface automatically translates the user's actions into commands for the system. 
This is in effect automatic code generation. 

This paradigm shift from commands to direct manipulation has led to the massive growth in computer usage worldwide, and the emergence of smartphones 
and tablets has confirmed the advantage of the approach. 
Numerous applications, in all fields, are available and usable at the touch of a finger. 
Users no longer need to memorize command lines with (sometimes) complex syntax. 
They act directly and immediately see the results of their actions. 

Paradoxically, in the field of programming, developers continue to write
commands, i.e. lines of code. 
Of course, they use sophisticated development environments (IDEs) that simplify their tasks (from assisting in
writing code to "refactoring" entire sections of code). 
For a professional computer scientist, this is not a problem. 
The same is not true for a novice programmer \cite{cheah2020factors}, \cite{kadar2021study},
\cite{javier2021understanding}. Indeed, writing lines of code involves imagining the effect those lines would produce when executed.

Our project asks the following question: can we develop a system that 
allows programming by seeing and manipulating the effect of the code we aim to produce? 
In fact, is it possible to apply the principle of direct manipulation, illustrated above, to the activity of programming, 
especially for novice programmers? 
Indeed, a beginner often has difficulty finding an algorithm, i.e., a method that achieves a desired result. 
However, being able to directly manipulate the data displayed on the screen, carry out successive tests, backtrack,
if necessary, and see the results obtained directly in the data provides significant help
\cite{naps2002exploring}, \cite{hundhausen2002meta}, \cite{marwan2020adaptive}. 
This way of proceeding, as recommended by Bret Victor, is easier for a beginner to access than constructing 
a mental representation of the algorithm being implemented \cite{victor2012learnable}.

AlgoTouch was designed to answer the two previous questions. 
It is a system that creates programs by directly manipulating the data that they operate on, mainly variables and arrays.

In this article, we begin by presenting the context of our work. 
We continue by describing the process of generating code from user manipulations. 
This second section details the machine model, the data manipulated, the permitted operations, 
the actions enabling these operations, and the associated code. Particular attention is paid to the implementation of conditionals and iterations. 
Finally, we take stock of our work before concluding with a discussion of future possibilities. 

\section[Programming by direct manipulation]{Programming by direct manipulation}
As mentioned earlier, direct manipulation has become common practice for computer users
(in the broad sense), but not very widespread in the field of code production.
In this section we discuss the main systems for
code production  by direct manipulation of data, and extract their main characteristics to motivate the objectives for the design of AlgoTouch.

\subsection[Related works]{Related works}
\label{sec:related_works}
Although the earliest work concerning programming by data manipulation is old (1975), to our knowledge relatively few systems have been proposed  in the intervening half-century.
This is quite paradoxical, as pointed out in \cite{luxton2018introductory}:
``{\em Given the increasing recognition of the importance of computational thinking in higher education, it is expected that
more developers will focus on building and evaluating tools in this area.}''

It is not our intention to list all the work in this area. 
Rather, we only consider work that deals with direct manipulation for the purpose of 
creating programs by novice programmers. 

Programming by direct data manipulation falls under the category of visual programming.
However, it differs from the approaches used by software such as Scratch \cite{resnick2009scratch}, Alice \cite{cooper2000alice}, and Blockly \cite{trower2015blockly}, which simplify the creation of code by assembling graphical building blocks, thus avoiding syntax problems.
Indeed, programmers do not manipulate the program data but create the  sequence of program instructions as they would with a text editor. Weidman \cite{weidmann2022bridging} pointed out: ``{\em However, block-based programming languages are similar to textual programming in that they still require users to mentally convert verbal instructions to state updates.}''

The literature also mentions programming by example (PBE)
\cite{gulwani2016programming} and programming by demonstration (PBD) \cite{cypher1993watch}.
In PBE, the user specifies the expected results of a program based on input data and the system attempts to automatically infer the code. 
The user does not describe the operations (the algorithm) that transformed the data to obtain the result.
On the contrary, in the case of PBD, the user directly shows the system 
how to perform a task by performing the actions themselves (e.g., manipulating an object or interacting with an interface).
These actions are then recorded by the system to produce a program.
It is in this last specific aspect of PBD that our work falls.

The Pygmalion system \cite{smith1975pygmalion}, developed by Smith in 1975, is one of the first examples of
direct manipulation programming. 
This innovative system allowed users to visually manipulate data on a graphical screen using icons, 
thus creating programs intuitively.
Smith named this approach "Programming by Demonstration".
Pygmalion's operation was simple:
the user's manipulations were recorded as a program, which could then be
executed to reproduce the actions. 
The system was Turing-complete, including conditionals, loops, and recursion.
The handling of conditionals was particularly innovative: the code for the true case of the conditional was generated immediately, 
while the false case remained undefined until a data manipulation involving this other case. 
We will see later that AlgoTouch handles conditionals in a manner that is similar to Pygmalion. 

SmallStar, developed by Halbert \cite{halbert1984programming} in 1984, builds on and adapts Pygmalion's ideas
to the Xerox Star system. 
This system consists of two separate windows: 
one dedicated to the manipulation of objects (such as files and folders) and 
the other to displaying code, generated exclusively by these interactions.
Recording code in SmallStar is a controlled process.
The user must specify the beginning and end of the sequence to be recorded. 
By default, code is added in sequence after the last existing instruction. 
However, it is also possible to insert instructions before a specific line or
delete lines of code. 
The program can be executed in full or line by line, under the user's control.
The construction of control structures also relies on direct manipulations.
To define a conditional, the user first records the necessary actions, and then
selects and encapsulates them in a
predicate that determines the execution conditions.
Similarly, creating a loop requires recording the
actions to be repeated, selecting them, and asking the system to generate the loop. This loop can then iterate
over a set of data selected according to their characteristics.

The Emacs editor \cite{stallman1993gnu}, while initially designed for text editing, incorporates a rudimentary  approach
to programming by demonstration through its macro system. 
Repetitive tasks can be recorded in the form of macros, then replayed to automate text and data manipulation.
The Emacs macro system provides great flexibility. 
Macro code can be edited verbatim, allowing the addition of control statements, 
conditional structures, loops and even structuring the code with
functions and parameters.
The AlgoTouch system borrows the notion of macros from Emacs. 

Alvis Live! \cite{hundhausen2009can} is a visual programming system that allows users to view, in
real-time, the state of program data while instructions are being constructed.
The user can click on an instruction written in pseudo-code (called SALSA) and observe the state of the variables resulting from the execution up to that point. One of the original features is the visualization of indices on arrays. To this end, the system introduces the concept of "index", an integer variable specifically associated with an array. In the interface, the index points to the array cell corresponding to its value, thus facilitating understanding for the user.
Furthermore, the program can either be typed or generated by direct manipulation of the program data.
The available version of Alvis Live! limits code generation by manipulation to array traversal loops and 
simple decision structures without an \verb+else+ clause.
Also, the system does not allow Turing-complete code to be generated by direct manipulation of the data, 
which restricts its scope of application. 
Direct manipulation operations remain very limited and require numerous dialog windows. 
Alvis Live! is designed primarily as a real-time visualization system for the state of a running program. 
We will see in section \ref{sec:index} that AlgoTouch also uses the notion of indexes on arrays. 

CodeInk \cite{scott2014direct} is a direct manipulation tool for producing Python code by interacting with
data structures.
The interface is divided into two areas: one for user-manipulated objects, and
the other for the generated Python code.
CodeInk allows the user to manipulate variables, lists, and binary trees.
The supported operations are assignment, insertion and comparison.
Assignment and insertion operations generate executable Python code. 
CodeInk does not generate code for comparisons; these are simply noted as comments
As a result, it is not possible to create complex control structures such as
loops and conditionals, limiting the tool's ability to produce Turing-complete programs.
Furthermore, the CodeInk project appears to be unfinished.
Despite its limitations, CodeInk is an interesting example of
direct manipulation of data structures for code generation.
The tool offers an intuitive interface and allows users to visualize the link between 
the manipulations performed and the code generated, just as with AlgoTouch. 

ManipoSynth \cite{hempel2022maniposynth} is a functional programming system distinguished by its ``bimodal'' approach to code generation.
The user can create Caml programs by directly manipulating data or by writing code in text form. 
ManipoSynth is value-centered. 
The user manipulates data and values to produce the corresponding code. 
This approach allows for intuitive and visual program creation, as the user focuses on the desired results 
rather than the code syntax.
ManipoSynth is also non-linear. 
The user can leave parts of the code blank and return to them later.
This flexibility allows for more fluid and exploratory program creation, 
without worrying about the code structure from the outset.
 ManipoSynth offers the ability to synthesize code from examples of
expected results. 
This feature is particularly useful for beginners or for complex tasks where
it is difficult to visualize the code upfront. 
Because ManipoSynth is bimodal, the user can interact with both the code and the data. 
They can choose the modality that best suits their situation and preferences.
This flexibility makes ManipoSynth accessible to a wide range of users: from beginners
to programming experts. 
In short, ManipoSynth aims to be Turing-complete and non-linear, features also present in AlgoTouch.

AlgoT \cite{thorgeirsson2024comparing} is a programming learning system based on the manipulation of program data.
The AlgoT interface is divided into three main sections: 
the developed programs with their code, 
the program variables, 
and the available operations.
When creating a new program, the user can add operations which can then be used to build others.
Code generation in AlgoT is done instruction by instruction.
At each step, the user must determine the operation or calculation to be performed, based on the contents of the variables.
Gradually, the code is fleshed out, and the content of the variables is updated accordingly.
AlgoT produces the code in a specific algorithmic language, characterized by simple conditional structures, without \verb+else+ clause.
Note that the language does not support writing loops; rather, repetition is handled by recursion.
Once the program is fully written and therefore fully executed, the user can 
navigate through the code and view the state of variables at each line of execution.
In studies, Graf et al. \cite{graf2024assessing} evaluated  the benefits of live programming 
for program comprehension in higher education, highlighting the educational interest of this method.
Although based on the manipulation of
data, AlgoT's code generation approach is clearly different from that of AlgoTouch.

Other tools, such as Sketch-n-Sketch \cite{chugh2016programmatic}, \cite{hempel2019sketch} 
and Twoville \cite{johnson2023computational}
allow vector graphics to be manipulated and 
associated with corresponding code.
The user can interact either with the graphic or with the code, and visualize the changes made in real time.
These tools are also called bimodal systems because they allow simultaneous interaction on two distinct levels: 
the visual representation and the underlying code.
Unlike our approach, these systems do not focus on manipulating program data (variables), but exclusively on graphics. 
However, this work illustrates the interest of direct manipulation as an efficient method for code generation.

\subsection[Objectives]{Objectives} \label{sec:objectives}
Analysis of existing systems that use data manipulation to generate code, 
along with Bret Victor's thoughts \cite{victor2012learnable} on program design, led to the following key principles for developing AlgoTouch:
\begin{description}
\item [Value-centered.] Code is generated solely by manipulating the data. 
This approach allows users to focus on the results they want to achieve, rather than the syntax of the code
\cite{victor2012learnable}.

\item [Non-linearity.] Code can be generated partially on data values. 
Even if parts of the code are not complete, it can still be executed.
This flexibility allows for more fluid and exploratory program creation
 \cite{hempel2022maniposynth}, \cite{smith1975pygmalion}.
\item [Visibility of data and program state.] Users should always be presented with a
visual representation of program data. 
This makes it easier to understand how the program works and correct errors \cite{victor2012learnable}.
\item [Clear semantics of operations.] 
The system must allow for intuitive understanding of how the operations work 
\cite{victor2012learnable}.
\item [Reactivity.] Users must be able to execute the instructions produced to immediately see their
effect on the problem data. 
Such functionality should allow the user to move through the code
to execute one or more instructions, and to possibly correct the code based on the observed result 
\cite{victor2012learnable}.
\item [Turing-completeness.] AlgoTouch should allow generating Turing-complete code with data, conditionals and loops. 
This means that the tool must be able to generate any possible computer program, giving it great power 
and flexibility  \cite{mcguffin2020categories}.
\end{description}
To our knowledge, AlgoTouch is a unique tool in the sense that no other tool for producing code in an imperative language meets all these objectives.

\section[Code generation by direct manipulation]{Code generation by direct manipulation}
In this section, we are interested in the realization of Turing-complete programs constructed
only by direct manipulations of the user as the means of interaction. 
The framework of this project is inspired by the report of a working group of the French Academy of Sciences \cite{berry2013enseignement}, which describes the four major concepts of computer science: algorithm, machine, language, and information.
Indeed, our approach to programming consists first in finding the algorithm by directly manipulating 
the data of the problem. 
These manipulations and the associated data depend on the available operations 
of a simple machine whose model will be explained later. 
Once the algorithm has been discovered, and the data represented
in formats adapted to the machine, it must be transformed into a program so that it can run on the machine. 
It is the role of a programming language to describe the program.

In this section, we first specify the target machine model, the chosen graphical interface, and the programming language that supports the programs produced by AlgoTouch. With these in hand, we describe how the various user manipulations produce code. Finally, we present the different execution modes available in AlgoTouch.

\subsection[Machine Model]{Machine Model}
\label{sec:machine}
To effectively support programming learning, it is essential to define a pedagogical structure
that makes the underlying mechanisms of abstract machines explicit. 
According to the review article \cite{fincher2020notional}, ``notional machine'' refers to a 
pedagogical tool designed to facilitate the understanding of certain aspects of programs or 
programming.

AlgoTouch assumes a simple underlying stored-program Von Neuman machine that is readily explained to a novice user of the system. The machine consists of a memory, a processor, an input device, and an output device. Memory elements contain integers or characters, and are identified by integer addresses. I/O is direct between memory locations and the device. The processor’s arithmetic-logical unit supports the usual integer arithmetic operations and comparisons of memory locations. The processor’s control unit supports a single sequential thread of execution, conditional flow of control based on comparisons, and non-recursive procedure call and return.

\subsection[AlgoTouch Interface]{AlgoTouch Interface}
\label{sec:interface}
Since a program is a series of operations to modify the state of the machine's memory
for a specific purpose, the AlgoTouch user must be able to simulate it by manipulating the data
in memory using only processor operations.
AlgoTouch therefore contains an area displaying memory data in the form of icons.
Manipulations of this data using touch gestures with a mouse or finger on an Interactive Whiteboard (IWB) allow the following operations to be performed: 

\begin{description}
\item [Read/write in memory]: copy or modify the value of a variable or a cell of an array.
\item [Operation]: 
display the operator of the calculation unit, and then copy the data in memory into operand registers.
\item [Inputs/Outputs]: drop a ``keyboard'' icon on a piece of data to enter it or drop a
data on a ``screen'' icon to display it.
\end{description}

The interface, shown in Figure \ref{fig:interface}, is designed to allow simultaneous visualization of data,
available operations, and program code. Thus, AlgoTouch has been structured according to this logic, with several
separate areas:
\begin{figure}[htb]
\centering
\includegraphics[width=\textwidth]{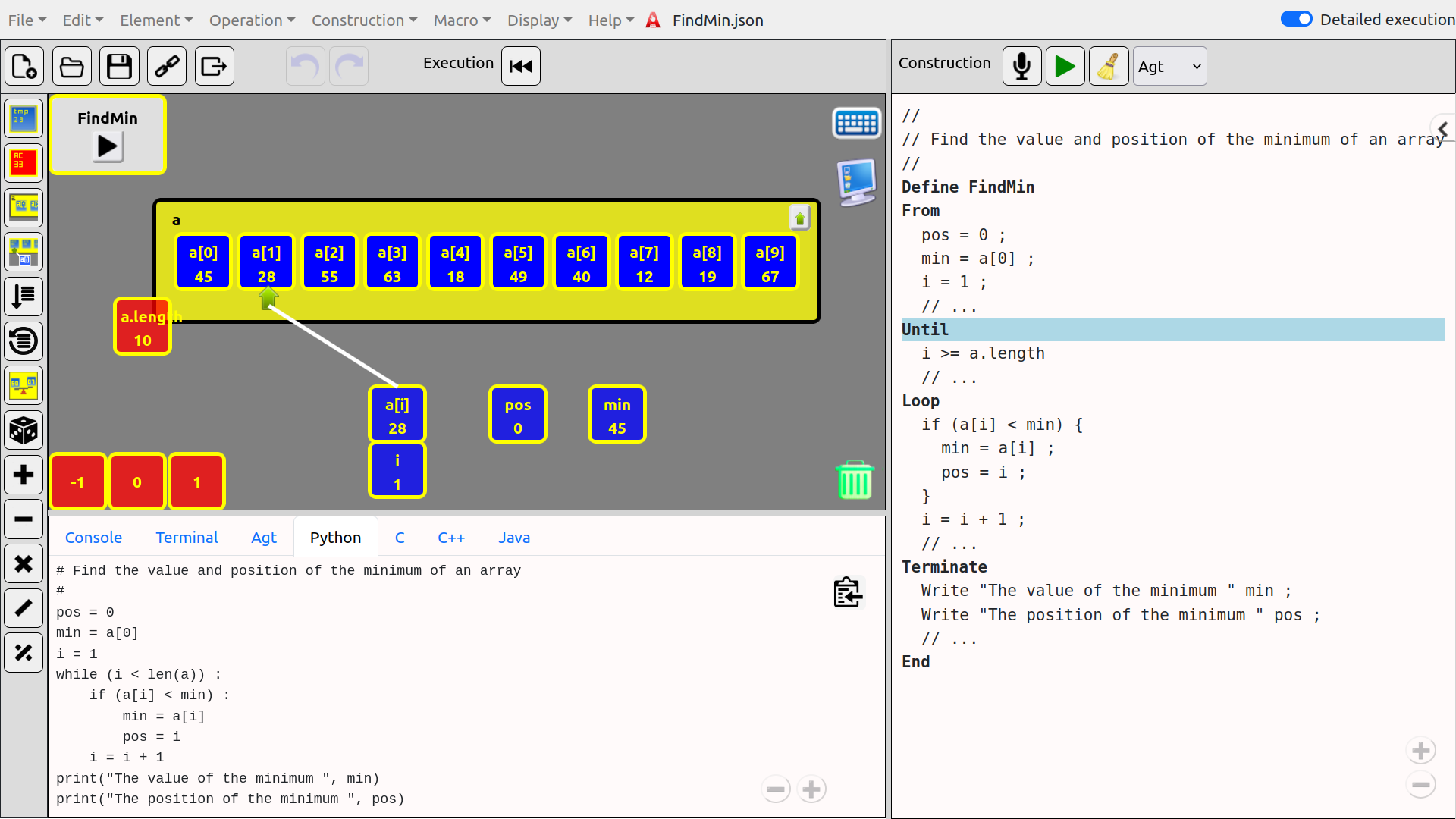}
\caption{ 
The AlgoTouch interface showing a program under construction, called {\tt FindMin}, for searching an array.
It displays the smallest value and its position.
The Workspace, in the center, presents an array {\tt a} of 10 elements indexed by the variable {\tt i}, and two variables {\tt pos} and {\tt min}.
The AGT program is shown on the right, and the Python version is available at the bottom.}
\Description{AGT Interface}
\label{fig:interface}
\end{figure}
\begin{enumerate}
\item \textbf{Workspace:} 
An area dedicated to manipulations, which contains the elements currently in use such as variables, 
arrays, associated indexes, as well as programs (macros), each represented 
as icons.
This area also includes common constants (-1, 0, 1) as well as icons for input and
output. 
It is a simplified representation of the elements contained in the machine's memory.

\item \textbf{Left Sidebar}: 
Allows the addition of elements such as variables, arrays, indexes, macros, or operators (comparator, arithmetic operators).

\item \textbf{Instructions Area}: 
Located on the right, this area displays the code of the program being built. 
The code is automatically generated and cannot be edited manually. 
It is, however, possible to select an instruction or a portion of the code. 
The display is provided in a language specific to AlgoTouch, called AGT, which is described later.
Other languages are available: Python, C, C++, and Java.

\item \textbf{Construction Area}: 
Located above the Instructions Area, it contains buttons to (1)
record manipulations as code, (2) execute parts of the program, (3) delete 
instructions and, (4) choose the construction language.

\item \textbf{Console and Export Area}: 
Located under the Workspace, with several tabs, it presents by default a console where the system displays the trace of the user's actions translated into lines of code
and the various system messages. 
The other tabs are used to display the translation of the macros into different languages,
with an option for exporting the code to an external tool.

\item \textbf{Toolbar}: 
Located above the Workspace, it offers common actions such as
creating new programs, opening or saving files, generating URLs, and exporting. 
A button also allows users to execute the program in animation mode, making visible both 
the executed instruction and its effect on the data.

\item \textbf{Menu Bar}: 
Located at the very top, it offers features already available graphically, as well as 
advanced functions such as code reorganization.
\end{enumerate}

Although AlgoTouch is a direct data-manipulation programming system in which the user does not write any line of code, the graphical interface contains an Instructions Area that displays the code generated in the AGT language.
This language is closely linked to the operating principles of AlgoTouch. It is a very simple Turing-complete language, whose syntax is inspired by C, Java, and Eiffel. 
Its syntax and semantics will be specified below together with presentation of the related data manipulations. 
Despite its simplicity, many programs associated with basic computer 
algorithms are feasible (e.g. searching and sorting in arrays by
example), as described in section \ref{sec:liste_prog}.
Note that an AGT program can be automatically transformed by AlgoTouch 
into any of the languages mentioned above.

The interface was designed respecting as much as possible the eight rules set out in \cite{shneiderman2010designing}:
\begin{enumerate}
\item Strive to make the interface consistent.
\item Allow shortcuts for common operations.
\item Provide quick and informative feedback.
\item Design dialogues that end with a clear action.
\item Propose simple ways to prevent and correct errors (robustness).
\item Allow reversibility of actions.
\item Promote user control.
\item Reduce short-term memory load.
\end{enumerate}
\subsection[The data]{The data}
In this section, we describe the different data that are available in AlgoTouch: 
variables, constants, arrays, and indexes,
as defined by the user with a name and a role, and  initialized. 
They are displayed on the Workspace and can be moved there.

\subsubsection[Variable ]{Variable }
A variable corresponds to a memory location that can be modified. 
It contains a value of the following two types: integer or character. 
\begin{figure}[htbp]
\centering
\begin{subfigure}[b]{0.49\textwidth}
\centering
\begin{minipage}[t]{0.30\textwidth}
\begin{tcolorbox}[colback=mygray, colframe=white, boxsep=0pt, left=0pt, right=0pt, top=\topboxpadding, bottom=\bottomboxpadding]
\centering
\includegraphics[scale=0.5]{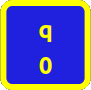}
\end{tcolorbox}
\end{minipage}
\end{subfigure}
\hfill
\begin{subfigure}[b]{0.49\textwidth}\centering
\begin{minipage}[t]{0.30\textwidth}
\begin{tcolorbox}[colback=mygray, colframe=white, boxsep=0pt, left=0pt, right=0pt, top=\topboxpadding, bottom=\bottomboxpadding]
\centering
\includegraphics[scale=0.5]{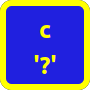}
\end{tcolorbox}
\end{minipage}
\end{subfigure}
\caption{An integer variable {\tt q} initialized to 0, and a character variable {\tt c} initialized to '?'.}
\label{fig:variables}
\Description{AGT Variables}
\end{figure}

Our work focuses on the production of code by manipulation of the
data, and does not require addressing other variable types: 
real, string, boolean, etc. 
A variable is represented by a blue square\footnote{The blue color indicates that the value can be modified by program.} with its name and value. Figure \ref{fig:variables} shows two variables of different types: integer and character.

\subsubsection[Named constant ]{Named constant }
\begin{figure}[htbp]
\centering
\begin{subfigure}[b]{0.49\textwidth}
\centering
\begin{minipage}[t]{0.30\textwidth}
\begin{tcolorbox}[colback=mygray, colframe=white, boxsep=0pt, left=0pt, right=0pt, top=\topboxpadding, bottom=\bottomboxpadding]
\centering
\includegraphics[scale=0.5]{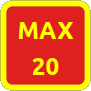}
\end{tcolorbox}
\end{minipage}
\end{subfigure}
\hfill
\begin{subfigure}[b]{0.49\textwidth}
\centering
\begin{minipage}[t]{0.30\textwidth}
\begin{tcolorbox}[colback=mygray, colframe=white, boxsep=0pt, left=0pt, right=0pt, top=\topboxpadding, bottom=\bottomboxpadding]
\centering
\includegraphics[scale=0.5]{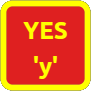}
\end{tcolorbox}
\end{minipage}
\end{subfigure}
\caption{An integer constant {\tt MAX} of value 20, and a character constant {\tt YES} of value 'y'.}
\label{fig:constants}
\Description{AGT Constants}
\end{figure}

A named constant corresponds to a memory location that cannot be modified. A named constant is represented by a red 
square\footnote{The red color indicates that the contents cannot be modified 
by a program. More generally, red is used to indicate prohibited actions 
to the user.} with its name and value. 
Figure \ref{fig:constants} shows two named constants of different types.

\subsubsection[Literal constant]{Literal constant}
A literal constant is a number or character defined by the programmer to perform assignments or
operations. It is represented by a red square containing only a value (no name). Literal constants are
present in the Workspace to facilitate their manipulation by drag and drop, like other AlgoTouch data. 
For example, to divide {\tt x} by 3, you must define 3 as a constant. By default, the
constants -1, 0 and 1, which are commonly used in programming, are predefined and present in the Workspace
(see Figure \ref{fig:interface}). 
This makes it easy to perform operations with these constants without having to use the keyboard, especially on tablets or IWBs.

\subsubsection[Array]{Array}
An array is an ordered sequence of variables. 
It has a name, and each variable in the array has that name followed 
by an index to specify the element.
For example, for an array {\tt a} of length 10, 
{\tt a[0]} is the first element, {\tt a[4]} is the fifth element, 
and {\tt a[9]} is the last element.

\begin{figure}[htbp]
\begin{tcolorbox}[colback=mygray, colframe=white, boxsep=0pt, left=\leftboxpadding, right=\rightboxpadding, top=\topboxpadding, bottom=\bottomboxpadding]
\centering
\includegraphics[width=\textwidth]{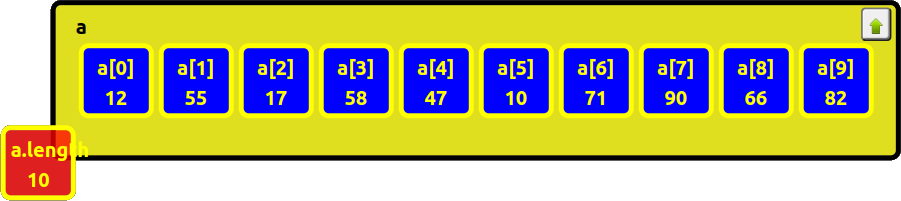}
\end{tcolorbox}
\caption{Representation of an array {\tt a} of 10 integers, and the associated constant, {\tt a.length},
representing its length.
Each cell in the array can be manipulated like any variable.}
\label{fig:array}
\Description{AGT Arrays}
\end{figure}

An AlgoTouch array is represented by a framed sequence of blue squares 
numbered from 0, as shown in Figure \ref{fig:array}. 
The length of an array {\tt a} is indicated by a constant (a red square) named
{\tt a.length}, which is displayed in the lower-left corner of the array.
The array and its length can be moved on the screen as a whole, while array cell and the array length can be manipulated as individual memory elements.
By default, when created, an array has a length of 10 and the
cells contain random values between 0 and 100. 
These characteristics can be modified.

\subsubsection[Index ]{Index }
\label{sec:index}
To locate an element of a sequence of numbers on a whiteboard, it is customary to use
one finger (usually the index finger). Similarly, in programming, an index identifies the position of an element in an array of
values. With AlgoTouch, as in ALVIS Live! \cite{hundhausen2009can} and
CodeInk \cite{scott2014direct}, an index is a specific variable. 

\begin{figure}[htbp]
\begin{tcolorbox}[colback=mygray, colframe=white, boxsep=0pt, left=\leftboxpadding, right=\rightboxpadding, top=\topboxpadding, bottom=\bottomboxpadding]
\centering
\includegraphics[width=\textwidth]{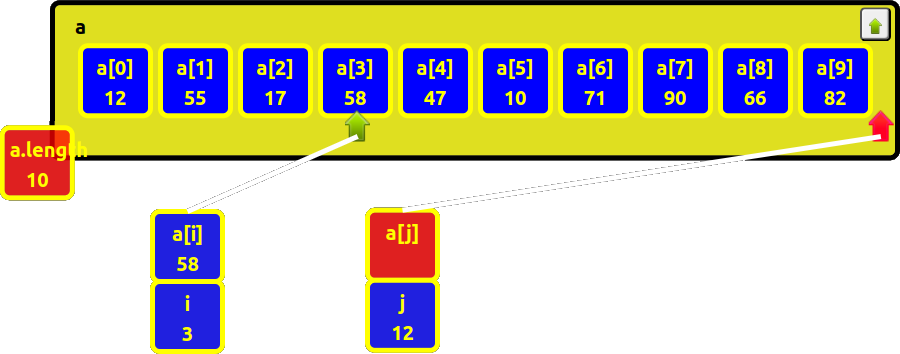}
\end{tcolorbox}
\caption{Array {\tt a} of 10 elements with two indexes, {\tt i} and {\tt j}.
Index {\tt i} is 3 and designates the value 58 in the array. 
Because index {\tt j}, which is 12, is outside the array the index does not designate any cell of the array, and
it is forbidden to use {\tt a[j]}. The button in the top right corner 
containing a green arrow is used to quickly define an index 
for the array {\tt a}.}
\label{fig:array2}
\Description{AGT Indexes}
\end{figure}

In the AGT language, an index is a particular variable, of type index, associated with one and only one array. For example, in 
Figure \ref{fig:array2}, the index {\tt i} is associated with the array {\tt a}. Like any variable, the index is represented by a blue square with its
value. 
When it is created, the value of an index is 0.
The display of an index contains two other visual aspects:
\begin{enumerate}
\item an arrow from the index (say {\tt i}) to the corresponding cell of the array ({\tt a[i]}). 
This facilitates a visual understanding of the notion of index and the associated cell of the array, and is identical to the technique used in Alvis Live!.
\item The blue square, above the variable {\tt i}, represents the element 
{\tt a[i]} with its value, 58, as illustrated in Figure \ref{fig:array2}. 
The presence of the indexed box {\tt a[i]} allows the user to manipulate 
it directly, without using a dialog box, as in Alvis Live!.
The value is more concrete because it is directly visible and can be 
manipulated like any other variable. 
When the index value is outside the limits of the array, the color of the square is red, 
indicating that it is impossible to consult or modify the value of the non-existent array element. 

\end{enumerate}

\subsubsection[Data Declarations in AGT Code]{Data Declarations in AGT Code}
Visual creation of date generates corresponding AGT statements. 
As in most programming languages, variables are declared with their type. 
The following declarations recapitulate the examples from the previous paragraphs:
\begin{tcolorbox}[colframe=black, colback=white, boxrule=0.5pt]
\begin{Verbatim}[samepage=true]
int q ;  // integer variable
char c ; // character type variable
    
const int MAX = 20 ;
// constant of integer type and value 20
    
const char YES = 121;
// character type constant with value 121
// ascii code for the character 'y'
    
int a[10] = {12,55,17,58,47,10,71,90,66,82};
// array of integers of size 10
    
index i of a ; // index of array a
\end{Verbatim}
\end{tcolorbox}

Note that no declaration is associated with literal constants.

\subsubsection[Notion of global data]{Notion of global data}
All data in an AlgoTouch program, variables, constants, arrays and indexes, 
can be used at any time.
In the field of programming languages, we speak of ``global variables'' 
to distinguish them from ``local variables''. 
As noted in recent literature \cite{szydlowska2022python}, 
\cite{zeevaarders2021exploring}, these notions are abstract and 
difficult to understand. 
With AlgoTouch, all variables are defined at program launch, 
which is consistent with the machine model presented previously 
in section \ref{sec:machine}. 
This does not prevent the creation of numerous programs, 
as we will see in section \ref{sec:liste_prog}.

\subsection[Operations, manipulations, and code generation] {Operations, manipulations, and code generation}
As specified in section \ref{sec:machine}, AlgoTouch is based 
on a simple machine model, inspired by the operation of a computer, 
and focuses on a limited but fundamental set of operations: viewing and modifying data, performing calculations, comparing values, and managing inputs and outputs.
AlgoTouch must allow these operations to be implied by direct manipulation 
of the data. 
All interactions are designed to be intuitive, limited to selection (click) and
drag-and-drop actions, usable with a mouse on a normal screen or 
with fingers on a touch screen. 
In addition, each action must be automatically transcribed into the 
corresponding code, ensuring consistency between visual manipulations and 
their translation into the programming language. 
For this purpose, during the manipulations, the code produced is displayed in the
Console Area. 
In the following, we detail how to perform these operations by manipulation to produce the associated code.

\subsubsection[Assignment ]{Assignment }
Assignment is the operation that replaces the contents of a data item with another value. The manipulation consists of 
dragging the value of the new data onto the data item to which it is to be assigned.
For example, to assign the value of the variable
{\tt v} to {\tt a[0]}, drag the
blue square corresponding to the variable {\tt v} onto the square corresponding to {\tt a[0]}.
Visually, to indicate that the assignment is possible, the frame of the square
{\tt a[0]} turns green. When the mouse is released, the value of
{\tt v} is assigned to {\tt a[0]}. The result of the assignment
is visible in the Workspace, and the associated instruction is produced in the Console Area.

\begin{figure}[htbp]
\centering
\begin{minipage}[b]{0.49\textwidth}
\centering
\begin{minipage}[b]{0.50\textwidth}
\begin{tcolorbox}[colback=mygray, colframe=white, boxsep=0pt, left=0pt, right=0pt, top=4pt, bottom=4pt]
\centering
\includegraphics[scale=0.4]{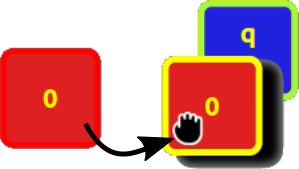}
\end{tcolorbox}
\end{minipage}
\caption{The constant {\tt 0} is moved onto the variable {\tt q}, producing the assignment {\tt q = 0}.}
\label{fig:assignment1}
\end{minipage}
\hfill
\begin{minipage}[b]{0.49\textwidth}
\centering
\begin{minipage}[b]{0.50\textwidth}
\begin{tcolorbox}[colback=mygray, colframe=white, boxsep=0pt, left=0pt, right=0pt, top=4pt, bottom=4pt]
\centering
\includegraphics[scale=0.4]{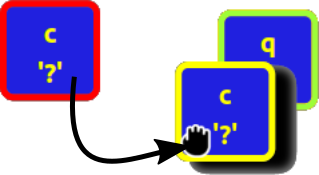}
\end{tcolorbox}
\end{minipage}
\caption{The variable {\tt c} is moved onto the variable {\tt q}, producing the assignment {\tt q = c}.}
\label{fig:assignment2}
\end{minipage}
\vfill
\Description{AGT Assignment}
\end{figure}

\begin{figure}[htbp]
\centering
\begin{minipage}[t]{0.30\textwidth}
\begin{tcolorbox}[colback=mygray, colframe=white, boxsep=0pt, left=0pt, right=-0pt, top=\topboxpadding, bottom=\bottomboxpadding]
\centering
\includegraphics[width=\textwidth]{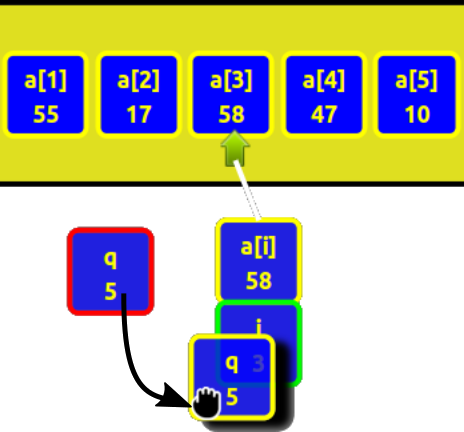}
\end{tcolorbox}
\caption{The variable {\tt q} is moved onto the index {\tt i} to produce the assignment {\tt i~=~q}. Thus,
{\tt a[i]} will change to 10, the value of {\tt a[5]}.\\}
\label{fig:assignment3}
\end{minipage}
\hfill
\begin{minipage}[t]{0.30\textwidth}
\begin{tcolorbox}[colback=mygray, colframe=white, boxsep=0pt, left=0pt, right=0pt, top=\topboxpadding,
bottom=18pt]
\centering
\includegraphics[width=\textwidth]{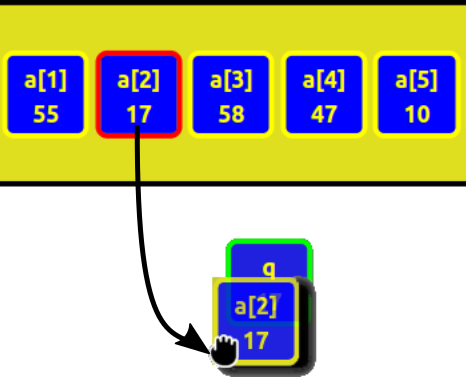}
\end{tcolorbox}
\caption{The cell {\tt a[2]} is moved onto the variable {\tt q} to produce the assignment {\tt q~=~a[2]}, the value of {\tt q} will then be 17.}
\label{fig:assignment4}
\end{minipage}
\hfill
\begin{minipage}[t]{0.30\textwidth}
\begin{tcolorbox}[colback=mygray, colframe=white, boxsep=0pt, left=0pt, right=0pt, top=\topboxpadding, bottom=10pt]
\centering
\includegraphics[width=\textwidth]{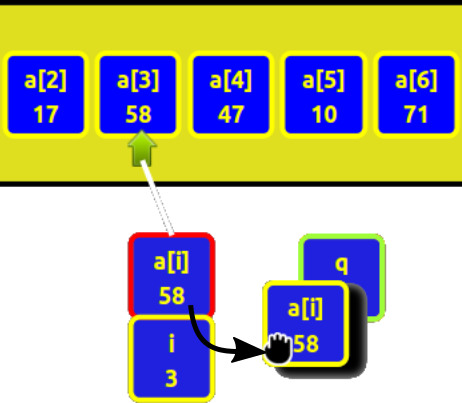}
\end{tcolorbox}

\caption{The variable {\tt a[i]} is moved onto the variable {\tt q} to produce the assignment {\tt q~=~a[i]}.
Thus, {\tt q} will take the value of {\tt a[3]}, i.e. 58.}
\label{fig:assignment5}
\end{minipage}
\Description{AGT Assignment}
\end{figure}

As in most common programming languages, the "{\tt =}" sign is the assignment symbol\footnote{It would have been better to use a different notation, but we chose to stick to the most commonly encountered syntax.}. The following examples
show the different assignments that arise from manipulations presented in corresponding figures:

\begin{center}
  \begin{minipage}{0.8\linewidth}
  \raggedright
  \texttt{q = 0;    // (Figure \ref{fig:assignment1})}

  \texttt{q = c ;   // (Figure \ref{fig:assignment2})}

  \texttt{i = q ;   // (Figure \ref{fig:assignment3})}

  \texttt{q = a[2]; // (Figure \ref{fig:assignment4})}

  \texttt{q = a[i]; // (Figure \ref{fig:assignment5})}

  \texttt{a[3] = q ;}

  \texttt{a[i] = a[j] ;}
  \end{minipage}
\end{center}


\subsubsection[Operations]{Operations}
The machine model described in section \ref{sec:machine} is equipped with a mechanism for performing
addition, subtraction, multiplication, integer division, and calculation of the remainder of an integer division.

\begin{figure}[htbp]
\centering
\begin{minipage}[b]{0.49\textwidth}
\centering
\begin{minipage}[b]{0.80\textwidth}
\begin{tcolorbox}[colback=mygray, colframe=white, boxsep=0pt, left=0pt, right=0pt, top=\topboxpadding, bottom=\bottomboxpadding]
\centering
\includegraphics[scale=0.4]{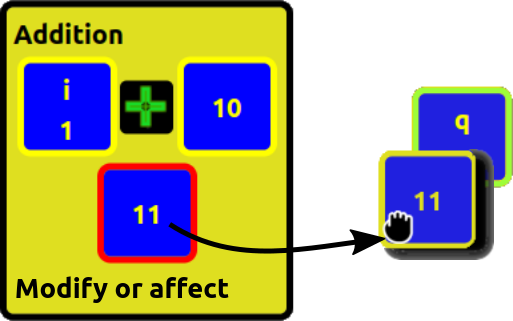}
\end{tcolorbox}
\end{minipage}
\caption{The result, of adding 1 ({\tt i}) and 10, which is 11, is moved onto the variable {\tt q} to produce
the assignment {\tt q~=~i~+~10}.}
\label{fig:affectation6}
\end{minipage}
\hfill
\begin{minipage}[b]{0.49\textwidth}
\centering
\begin{minipage}[b]{0.80\textwidth}
\begin{tcolorbox}[colback=mygray, colframe=white, boxsep=0pt, left=0pt, right=0pt, top=\topboxpadding, bottom=\bottomboxpadding]
\centering
\includegraphics[scale=0.37]{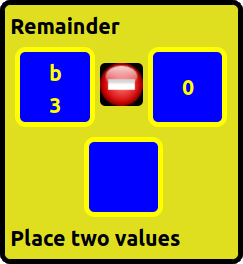}
\end{tcolorbox}
\end{minipage}
\caption{ Division by zero is prohibited. A prohibition sign replaces the operator,
and it is impossible to move the result.}
\label{fig:affectation7}
\end{minipage}
\vfill
\Description{AGT Calculation}
\end{figure}

To perform a calculation, the user first selects the operation, whereupon a small 3-slot window appears, as shown in the Figure
\ref{fig:affectation6}. After dragging the data from the two operands to the first two slots, the result
is displayed in the lower box.
In the same manner as assignment, all that remains is to drag the result into a data item,
whereupon the AGT code is produced. 
This mechanism is modeled on the calculation performed in an
arithmetic-logic unit: reading in memory of the two values to be processed,
calculating the result, and writing the result into memory.

Figure \ref{fig:affectation6} illustrates the addition operation, 
{\tt q~=~i~+~10} and Figure \ref{fig:affectation7} (an error case)
division by zero.
Operators are only binary. 
This limitation is not restrictive because every calculation is decomposed 
into a sequence of elementary 
calculations\footnote{It would be possible in principle to change the system 
so that one could use the result of a calculation as an operand of another 
operation to perform a more complex calculation. }.

Increment, i.e. adding 1, and decrement, 
i.e. subtracting 1, are common operations in programming. 
These two operations can be performed using an adder or a subtractor,
but AlgoTouch also offers direct manipulation: 
increment can be done by sliding the mouse (or the finger) 
starting from the left of the data to be incremented, up to it. 
To indicate that the increment is possible the border of the data square 
turns green (Figure \ref{fig:increment}). 
A decrement is carried out with a slide from right to left 
(Figure \ref{fig:decrement}).

These manipulations were introduced to simplify the use of indexes for array traversal. Sweeping the index to the right moves to the next cell to the right.
The direction of the gesture is consistent with the representation of arrays 
where index 0 is at the left end. 
This manipulation gives the user the impression of moving 
the arrow to the right.  
The result of these operations produces the following code:
\begin{Verbatim}[samepage=true]
        x = x + 1 ;
        i = i - 1 ;
\end{Verbatim}
\begin{figure}[htbp]
\centering
\begin{minipage}[b]{0.49\textwidth}
\centering
\begin{minipage}[b]{0.80\textwidth}
\begin{tcolorbox}[colback=mygray, colframe=white, boxsep=0pt, left=0pt, right=0pt, top=50pt, bottom=\bottomboxpadding, height=140px]
\centering
\includegraphics[scale=0.4]{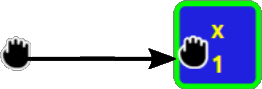}
\end{tcolorbox}
\end{minipage}
\caption{ Increment by rightward sweep to {\tt x}, which then becomes 2.\\}
\label{fig:increment}
\end{minipage}
\hfill
\begin{minipage}[b]{0.49\textwidth}
\centering
\begin{minipage}[b]{0.744\textwidth}
\begin{tcolorbox}[colback=mygray, colframe=white, boxsep=0pt, left=0pt, right=0pt, top=\topboxpadding, bottom=\bottomboxpadding, height=140px]
\centering
\includegraphics[scale=0.407]{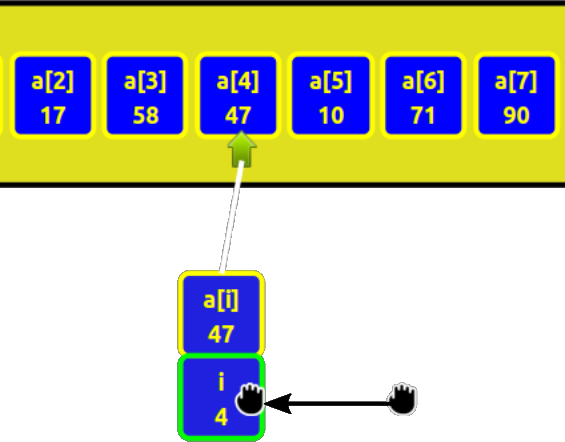}
\end{tcolorbox}
\end{minipage}
\caption{Decrement by leftward sweep to {\tt i}, which then becomes 3. The value of {\tt a[i]} then becomes 58.}
\label{fig:decrement}
\end{minipage}
\vfill
\Description{AGT Conditonal}
\end{figure}

\subsubsection{Comparisons}
\label{sec:comparisons}
\begin{figure}[ht]
\centering
\begin{minipage}[b]{0.33\textwidth}
\begin{tcolorbox}[colback=mygray, colframe=white, boxsep=0pt, left=0pt, right=0pt, top=\topboxpadding, bottom=\bottomboxpadding]
\centering
\includegraphics[scale=0.35]{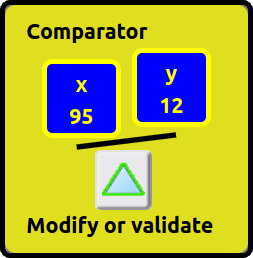}
\end{tcolorbox}
\end{minipage}
\caption{Representation of a comparator. The larger value tips the scale.}
\label{fig:scale}
\Description{AGT Conditional}
\end{figure}

The comparator of AlgoTouch’s underlying machine model can test data values and choose which instructions to execute next based on the result. 
It is represented as a two-pan scales, as shown in Figure \ref{fig:scale}. 
The user places a data item on each pan, and the scale tips to one side or the other, or remains balanced if the values are equal.

When users just manipulate data, they do not need a comparator 
because they can see the values, and do the comparison in their head.
An optional visualization mode called “blind mode” hides values, which forces users to adopt the computer’s perspective.
In blind mode, the comparator result plays a key role for deciding on future manipulations.
Blind mode is not detailed in this paper; its use and interest are developed in \cite{frison-hal-01753119} where it is presented as a tool that facilitates the transition between an "unplugged" approach and computer programming.
In either visualization mode, a comparator is essential for defining conditionals when building a program, as we will see in section \ref{sec:conditional}.

\subsubsection[Reading a value]{Reading a value}
\begin{figure}[htbp]
\centering
\begin{minipage}[t]{0.49\textwidth}
\centering
\begin{minipage}[t]{0.60\textwidth}
\begin{tcolorbox}[colback=mygray, colframe=white, boxsep=0pt, left=0pt, right=0pt, top=\topboxpadding, bottom=\topboxpadding, height=50px]
\centering
\includegraphics[scale=0.4]{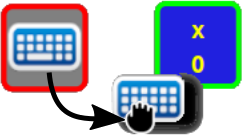}
\end{tcolorbox}
\end{minipage}
\caption{
Inputting a value into variable {\tt x} produces the instruction {\tt Read "Read~x~"~x}. }
\label{fig:read}
\end{minipage}
\hfill
\begin{minipage}[t]{0.49\textwidth}
\centering
\begin{minipage}[t]{0.60\textwidth}
\begin{tcolorbox}[colback=mygray, colframe=white, boxsep=0pt, left=0pt, right=0pt, top=\topboxpadding, bottom=\bottomboxpadding, height=50px]
\centering
\includegraphics[scale=0.4]{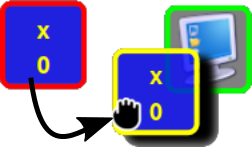}
\end{tcolorbox}
\end{minipage}
\caption{
Writing the value of variable {\tt x}
produces the instruction {\tt Write~"Value~of~x~"~x}.}
\label{fig:write}
\end{minipage}
\vfill\Description{AGT Write}
\end{figure}

Input of user data is performed by dragging the keyboard icon and dropping it onto a target variable (see Figure \ref{fig:read}). 
Hovering over the variable turns its border green to indicate that the assignment is permitted. 
A dialog window then opens containing a message for the user, and a field in which to enter the data value.
By default, the message for some variable, say {\tt x}, is {\tt "Read x"}, which can be changed to suit the problem at hand.

Changing the value of a variable via input is seen as the transfer 
of the value entered on the keyboard to the variable, and produces the code: 
\begin{Verbatim}[samepage=true]
        Read "Read x " x ;
\end{Verbatim}
In contrast, changing the value of a variable by directly editing the value in the variable’s property sheet is an incidental change that produces no code. 

\subsubsection[Writing a value]{Writing a value}
Output of a variable’s value is invoked by dragging the variable,
and dropping it onto the screen icon (see Figure \ref{fig:write}). 
Hovering over the screen icon turns the outline green.
A dialog window then opens with a message for the user, and the value of the variable.
By default, the message for some variable, say {\tt x}, is: 
{\tt "Value of x"}, which can be changed to suit the problem at hand.

Outputting the value of the variable is seen as transferring a copy of the value to the dialog window, and produces the code: 
\begin{Verbatim}[samepage=true]
        Write "Value of x " x ;
\end{Verbatim}

Understanding the need to explicitly output a value requires a small conceptual leap for a novice given that during direct manipulation the variable’s value is already visible in the Workspace. Specifically, the novice must learn the notion of “run time”, and understand that the Workspace is not necessarily kept up to date during execution, but that executing the generated output instruction will emit the value regardless.

\subsection[AlgoTouch program - macro concept]{AlgoTouch program - macro concept}
\begin{figure}[ht]
\centering
\begin{minipage}[b]{0.33\textwidth}
\centering
\begin{minipage}[b]{0.60\textwidth}
\begin{tcolorbox}[colback=mygray, colframe=white, boxsep=0pt, left=0pt, right=0pt, top=\topboxpadding, bottom=\bottomboxpadding]
\centering
\includegraphics[scale=0.4]{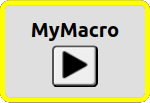}
\end{tcolorbox}
\end{minipage}
\end{minipage}
\caption{Macro AlgoTouch named MyMacro.}
\label{fig:macro}
\Description{AGT Macro}
\end{figure}

As we have seen, a user can interact directly with data to perform desired processing, and (as a side effect) AlgoTouch will produce corresponding instructions that are displayed in the Console Area.
Once a user has determined what actions to perform on the data, and after experimenting with them, they may be ready to keep the instructions in an appropriate order within a program, which can then be tested on different cases, and modified, if necessary. 
To do this, the user must activate Recording mode before performing the actions on the data. In Recording mode, generated instructions are inserted into a structure called a Macro. 
Each macro is displayed in the Workspace by an icon containing its name, and a black triangular button that can be used to invoke its code, as shown in Figure \ref{fig:macro}. Clicking the background of a macro icon selects its code for display in the Instruction Area on the right side of the screen.

Macros are AlgoTouch’s notion of subroutine—with their well-known properties.
AlgoTouch only defines macros of two types: 
Simple Macro
\footnote{The terms Simple Macro and Loop Macro will always be written with
capital letters since they are concepts of the AGT language.} and Macro Loop. 
We first cover Simple Macros.
A discussion of Loop Macros is deferred until section \ref{sec:macro_loop}.

\paragraph[Simple Macro ]{Simple Macro:}
\label{sec:simpleMacro}
A Simple Macro consists of a name, a role, and a sequence of instructions 
(possibly empty) enclosed  by the keywords {\tt Do} and {\tt End}, 
subsequently called the {\tt Do} block. 
For example, creating the macro {\tt Average}, with the role "Computes the average of 
the integers {\tt a} and {\tt b} in {\tt avg}", defines the following macro, which is then presented in the Instructions 
Area:

\begin{tcolorbox}[colframe=black, colback=white, boxrule=0.5pt]
\begin{Verbatim}[samepage=true]
//
// Computes the average of the integers a and b in avg
//
Define Average
    Do
        // ...
    End
\end{Verbatim}
\end{tcolorbox}

\begin{figure}[ht]
\centering
\begin{minipage}[b]{0.33\textwidth}
\centering
\begin{minipage}[b]{0.60\textwidth}
\begin{tcolorbox}[colback=mygray, colframe=white, boxsep=0pt, left=0pt, right=0pt, top=\topboxpadding, bottom=\bottomboxpadding]
\centering
\includegraphics[scale=0.4]{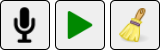}
\end{tcolorbox}
\end{minipage}
\end{minipage}
\caption{Buttons associated with macro construction : Record, Execute, Delete.}
\label{fig:construction}
\Description{AGT Macro}
\end{figure}

When created, a macro contains no instructions, only an empty comment 
({\tt // ...}).
To add instructions, the user selects the line {\tt Do}, and 
then activates Recording mode by clicking the Record button 
in the Construction Area (see figure \ref{fig:construction}). 
In Recording mode, the instructions produced by each data manipulation, in addition 
to being emitted to the Console, are also inserted line-by-line in sequence into 
the Do block, and the Recording button turns red to indicate that recording is active. 

For example, after recording, the code for the {\tt Average} macro might be:
\begin{tcolorbox}[colframe=black, colback=white, boxrule=0.5pt]
\begin{Verbatim}[samepage=true]
//
// Computes the average of the integers a and b in avg
//
Define Average
    Do
        s = a + b ;
        avg = s / 2 ;
    End
\end{Verbatim}
\end{tcolorbox}
\label{code:average}

After recording, the three buttons shown in Figure \ref{fig:construction} allow one to insert newly generated instructions, execute the code, or delete lines.
These functionalities will be illustrated in the following sections on conditionals \ref{sec:conditional} and iterations \ref{sec:loop}.
The execution triggered by the Execute Button will be referred to as execution in Construction mode, which is distinct from the standard execution mode of a macro.

\subsection[Conditional]{Conditional }
\label{sec:conditional}
Thus far, we have only discussed the generation of a straight-line program by direct data manipulation. The process is intuitive because the state of memory is updated synchronously with the development of the “code”. 
In contrast, the branch of a conditional presents a question: how to maintain alignment between the state of memory and the “code” being developed at a fork in the flow of control at a conditional branch? 

Upon reaching the test of a conditional, it is natural to proceed along the branch that matches the current memory state, but subsequent “coding” of the other branch requires (a) that all memory-state changes made along the first-coded branch be reverted, and requires (b) that the reverted memory state be altered so that it comports with the test taking on the opposite truth value. Only Pygmalion \cite{smith1975pygmalion}, Maniposynth \cite{hempel2022maniposynth} and AlgoTouch seem to have fully addressed this question.

A conditional is introduced at a given control point by defining a comparator, as described in section \ref{sec:comparisons}. The system assumes that when a conditional is inserted in a given data state, control is intended to proceed along the true branch. Accordingly, the only choices offered for the comparator’s relational test are those that would test true in that state. Although either branch of the conditional can be coded next, it is most convenient to code the true branch first because it corresponds to the data state that was just used to generate the relational test. For a similar reason, after the true branch has been coded, it is most convenient to continue coding after the conditional, deferring consideration of the false branch until later. The net effect then of having introduced the conditional is the interposition of a true test in what is otherwise straight-line code. Coding order is thus driven by data, as called for by the value-centered principle stated in section \ref{sec:objectives}.

When the user is ready to address the as-yet-uncoded else branch, they establish program input values such that when control reaches the conditional upon execution on Construction mode, the condition will be false. Control will then pass to the false branch, whereupon execution stops with a warning message stating that code is missing. Coding the else branch by direct manipulation then proceeds in the appropriate data state. 

We illustrate with a very simple program fragment that modifies variable {\tt v} by incrementing it when it is negative or zero, and decrementing it when it is positive. The user begins by assigning a negative value to {\tt v} (with Recording mode off), and then (with Recording mode on) performs the necessary manipulation to increment {\tt v}. The user then defines a new case by choosing a positive value for {\tt v} (with Recording mode off), and restarts execution in Construction mode. The system pauses where code for the positive case is missing, whereupon manipulations are performed (with Recording mode on) to decrement {\tt v}. We further detail the example below. 

In the first case, variable {\tt v} contains a negative value (-3). The user starts recording, and introduces a comparator from the Left Sidebar. The sign of {\tt v} is tested by dropping the values of {\tt v} and {\tt 0} onto the scales, and then activates the comparison by clicking the fulcrum button. AlgoTouch then lists the three comparisons that would test true in the given data state, as illustrated in the Figure \ref{fig:comparisons}: 
\begin{figure}[htbp]
\centering
\begin{subfigure}[b]{0.40\textwidth}
\begin{tcolorbox}[colback=mygray, colframe=white, boxsep=0pt, left=0pt, right=0pt, top=13pt, bottom=\bottomboxpadding, height=127px]
\centering
\includegraphics[scale=0.4]{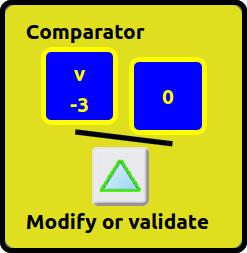}
\end{tcolorbox}
\end{subfigure}
\hfill
\begin{subfigure}[b]{0.59\textwidth}
\begin{tcolorbox}[colback=mygray, colframe=white, boxsep=0pt, left=0pt, right=0pt, top=1pt, bottom=0pt, height=127px]
\centering
\includegraphics[scale=0.315]{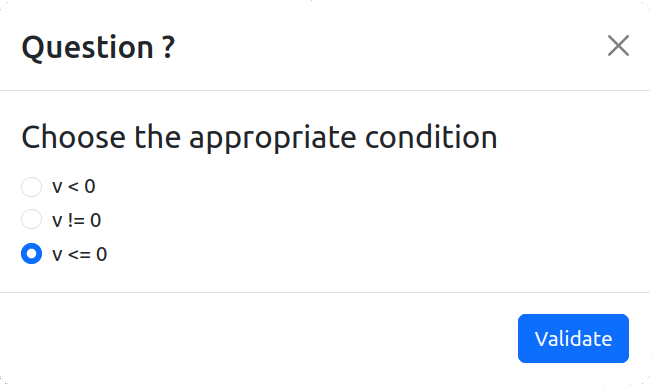}
\end{tcolorbox}
\end{subfigure}
\caption{ The comparator of {\tt v} and 0, and the three possible true relations in the given data state. 
}
\label{fig:comparisons}
\Description{AGT Comparisons}
\end{figure}

\begin{description}
\item [\textbullet]{\tt v~<~0}, because -3 is strictly less than 0,
\item [\textbullet]{\tt v~!=~0}, because -3 is different from 0,
\item [\textbullet]{\tt v~<=~0}, because -3 is also less than or equal to 0.
\end{description}

\begin{figure}[htbp]
\centering
\begin{minipage}[c]{0.49\textwidth}
\centering
\begin{minipage}[c]{0.60\textwidth}
\begin{tcolorbox}[colback=mygray, colframe=white, boxsep=0pt,
                  left=0pt, right=0pt, top=\topboxpadding, bottom=\topboxpadding]
\centering
\includegraphics[scale=0.4]{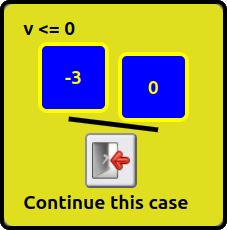}
\end{tcolorbox}

\end{minipage}
\caption{
Dialog box after selecting the condition {\tt v~<=~0} to be processed.
}
\label{fig:ifAfterSelection}
\end{minipage}
\hfill
\begin{minipage}[c]{0.49\textwidth}
\centering
\begin{minipage}[c]{0.60\textwidth}
\begin{tcolorbox}[colback=mygray, colframe=white, boxsep=0pt,
                  left=0pt, right=0pt, top=\topboxpadding, bottom=\bottomboxpadding,
                  height=100px, valign=center] 
\centering
\includegraphics[scale=0.4]{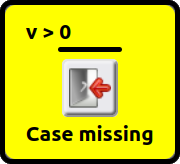}
\end{tcolorbox}
\end{minipage}
\caption{
Dialog box reporting that the case 
{\tt v~>~0} needs to be processed.
}
\label{fig:else}
\end{minipage}
\end{figure}

Of course, only the user knows which of the three tests is appropriate. 
When the user selects the third choice, {\tt v~<=~0}, the comparator’s dialog box changes to reflect the test that has been chosen, and provides a button for advancing the cursor beyond the conditional, as shown in Figure \ref{fig:ifAfterSelection}. 

The user then increments {\tt v}, which generates the statement {\tt v~=~v~+~1}, and results in the code: 
\begin{Verbatim}[samepage=true]
        if (v <= 0) {
            v = v + 1 ;
        } else {
            // v > 0
            // TO DO
        }
\end{Verbatim}
To continue coding after the conditional (unnecessary in this example), or to merely dismiss the dialog box, the user clicks ``Continue this case''. 

The comment ``{\tt //~TO~DO}'' in the else-clause indicates that the case {\tt v~>~0} has not yet been addressed. To complete the code, the user changes {\tt v} to a positive value, and then re-executes the conditional in Construction mode. At this point, AlgoTouch stops execution on the ``{\tt TO~DO}'' line, activates Recording mode, and displays a ``Case Missing'' dialog box (shown in Figure \ref{fig:else}). The user performs the manipulations that decrement {\tt v}, and dismisses the dialog box by clicking the ``Case missing'' button. 

The produced code is now complete: 
\begin{Verbatim}[samepage=true]
        if (v <= 0) {
            v = v + 1 ;
        } else {
            // v > 0
            v = v - 1 ;
        }
\end{Verbatim}
If no action were required for positive {\tt v}, the user could just close the ``Case missing'' dialog box, which would adjust the code to: 
\begin{Verbatim}[samepage=true]
        if (v <= 0) {
            v = v + 1 ;
        }
\end{Verbatim}

The interactive and visual approach to coding conditionals based on data guides the programmer in thinking about and implementing required behaviors, and conforms to principles set out by Bret Victor \cite{victor2012learnable}. 
Note that AlgoTouch does not allow compound conditions involving Boolean operators (AND, OR, NOT). Rather, in direct manipulation, comparisons are performed one after the other, which leads to nested conditionals. Thus, the absence of Boolean operators is not a limitation. 

\subsection[Loop ]{Loop }
\label{sec:loop}
To obtain a Turing-complete mechanism, all that remains is loops.
We chose to implement iteration in the form of a specific kind of macro called a Loop Macro. 

While conditional statements are allowed wherever simple instructions are permitted, loops are not.
Rather, each loop must be defined in a separate macro. 
By isolating the loop in a macro, the programmer can focus on its implementation. 
A loop in AlgoTouch is composed of several blocks, each built individually during development as described in the next section. 
In addition, once the macro is completely defined, its correct operation can be tested independently 
of the rest of the program. 
Finally, the user can use it in other parts of the program, especially to build nested loops, 
as explained in section \ref{sec:prog_complexes}.

\subsubsection[Loop Macro]{Loop Macro}
\label{sec:macro_loop}

Creating a loop structure through direct data manipulation is a 
challenging task. Indeed, it is not possible to simply record the 
sequence of actions to be performed, since by definition the number 
of iterations to execute is, in general, not known in advance. 
However, the programmer does know which operations must be repeated 
and what the stopping or continuation conditions of the loop are.
Programming languages generally offer several types of loop constructs 
for this purpose (such as {\tt do while}, {\tt repeat until}, {\tt for}, or {\tt while}).
AlgoTouch integrates a single generic loop structure that is sufficient for programming all algorithms. In addition, this saves users, especially beginners, from asking which type of loop to use. 
In the context of code generation through direct manipulation, we adopted a loop structure inspired by the Eiffel language \cite{meyer1992eiffel}, which clearly identifies the different components of a loop.

A Loop Macro consists of a name and four blocks: {\tt From},
{\tt Until}, {\tt Loop}, and {\tt Terminate}.
The user completes these blocks by direct manipulation, where {\tt From}, {\tt Loop}, and
{\tt Terminate} contain instructions (assignments, conditionals, inputs/outputs, and macro calls), and {\tt Until} contains a list of one or more conditions.  
After each step of block creation in Construction mode, it is possible to see the effect of execution on the data (the notion of {\bf reactivity}). 
This allows the validity of the program to be verified line by line or block by block.


For example, after creating the Loop Macro {\tt MyLoopMacro}, the following code is displayed in the Instructions Area. The blocks contain empty comments ({\tt // ...}) that will be gradually replaced by instructions. 
\begin{tcolorbox}[colframe=black, colback=white, boxrule=0.5pt]
\begin{Verbatim}[samepage=true]
    //
    // Example of a loop macro.
    //
    Define MyLoopMacro
        From
            // ...
        Until
            // ...
        Loop
            // ...
        Terminate
            // ...
    End
\end{Verbatim}
\end{tcolorbox}
\begin{itemize}
\item[\textbullet] The {\tt From} block contains preparatory instructions, 
including the initializations of the loop variables. 
It is therefore executed first and only once.
\item[\textbullet] The {\tt Until} block contains a list of exit conditions. 
Each condition is limited to a single comparison. 
This block is evaluated once after the {\tt From} block and before
the {\tt Loop} block, and then after each execution of the {\tt Loop} block. 
The exit conditions are evaluated sequentially. Once a condition is true, the following conditions are not evaluated,
and the program continues in the {\tt Terminate} block.
\item[\textbullet] The {\tt Loop} block is repeatedly executed so long as no exit condition has been met. .
\item[\textbullet] The {\tt Terminate} block contains instructions that are executed once 
after an exit condition has been met. 
\end{itemize}

\subsubsection{Data-aware programming}
AlgoTouch is designed to promote reasoning about all possible execution scenarios according to the data being processed.
For instance, when performing a sequential search for a value {\tt v} in an array {\tt a}, the value may either be present or absent.
In the first case, the goal generates the exit condition {\tt a[i] == v}, and in the second, {\tt i >= a.length}.
Both scenarios must therefore be anticipated through appropriate initialization and definition of program behavior.

In conventional languages, loops such as {\tt while} often rely on complex continuation conditions using Boolean operators, e.g.:
\begin{Verbatim}[samepage=true]
    while (i < a.length and a[i] != v)
\end{Verbatim}

In contrast, AlgoTouch defines loops through a list of simple exit conditions rather than a single Boolean expression:
\begin{Verbatim}[samepage=true]
    Until
        i >= a.length
        a[i] == v
\end{Verbatim}

This approach improves code readability and understandability.
By expressing exit conditions in implicit disjunctive normal form, it also removes the need for users to reason about Boolean algebra, thereby simplifying the design of control flow.
\subsubsection[Methodology for constructing a Loop Macro]{Methodology for constructing a Loop Macro}
\label{sec:method}
Building a loop by direct data manipulation in a visual environment requires an appropriate 
systematic methodology.
Indeed, although building the straight-line code of a Simple Macro by recording a linear sequence
of corresponding basic operations is obvious and without a plausible alternative, 
the same is not true for a Loop Macro.  

In keeping with the principle of non-linearity stated in section \ref{sec:objectives}, 
we propose a preferred creation order for the blocks of a Loop Macro that takes advantage 
of the ability to execute incomplete code, and that stems from our belief 
that algorithm development should progress 
from the general case to special cases.
Specifically, the user is encouraged to elaborate the blocks in the following order: the body 
({\tt Loop} block), the iteration-stopping criteria ({\tt Until} block), and finally the termination ({\tt Terminate} block). 
The initialization ({\tt From} block), on the other hand, can be generated at the beginning or completed progressively during the construction process.

The objective is to progressively define the algorithm from the general case to the particular 
cases: at each stage, AlgoTouch will be used, in application of the principle of {\bf non-linearity}.
We illustrate the benefits of the proposed elaboration order in the next section.

\subsubsection[Example of constructing a loop]{Example of constructing a loop}

\label{sec:loop_construction}
We illustrate code generation for a Loop Macro using as an example the inner loop 
of a standard implementation of Insertion Sort:  
given that array segment {\tt a[0..i-1]} is ordered, rearrange {\tt a[0..i]} so that it is ordered.  

We can imagine several alternative approaches: 
\begin{itemize}
\item[\textbullet] 
Make a copy of the value to be placed; scan from left to right to find where it is to be placed; 
shift the succeeding values to the right; copy the preserved value to its proper place  (version 0); 

\item[\textbullet] 
Move the value to be inserted gradually to the left until it reaches its final position (version 1);
\item[\textbullet] 
Make a copy of the value to be placed; scan the values from right to left, moving the larger elements to the right; copy the preserved value to its proper place (version 2). 
\end{itemize}

We adopt version 1, using it to illustrate how data-driven code generation determines 
coding order, and how this process follows the principle of non-linearity, 
as stated in section \ref{sec:objectives}. 

First, we address the general case, i.e., insertion into the sorted part.  Then, we consider the two special cases: insertion at the end of the sorted part, and insertion at the beginning of the array. 

\begin{figure}[htbp]
\centering
\begin{minipage}[t]{0.49\textwidth}
\centering
\begin{tcolorbox}[width=0.75\textwidth, colback=mygray, colframe=white, boxsep=0pt, left=8pt, right=0pt, top=\topboxpadding, bottom=\bottomboxpadding]
\centering
       
\includegraphics[scale=0.25]{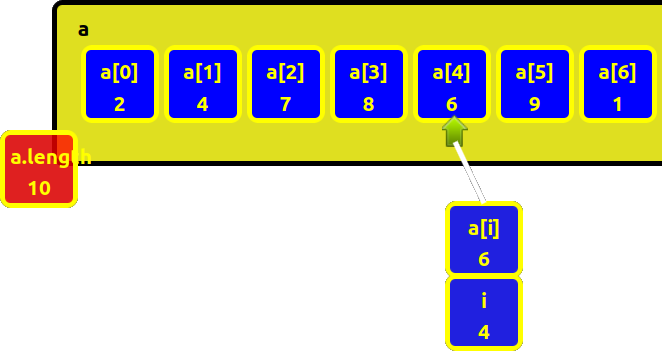}
\end{tcolorbox}
\caption{the value 6 is to be inserted between 4 and 7.}
\label{fig:insertion1}
\end{minipage}
\hfill
\begin{minipage}[t]{0.49\textwidth}
\centering
\begin{tcolorbox}[width=0.75\textwidth, colback=mygray, colframe=white, boxsep=0pt, left=7pt, right=0pt, top=\topboxpadding, bottom=\bottomboxpadding]
\centering
\includegraphics[scale=0.25]{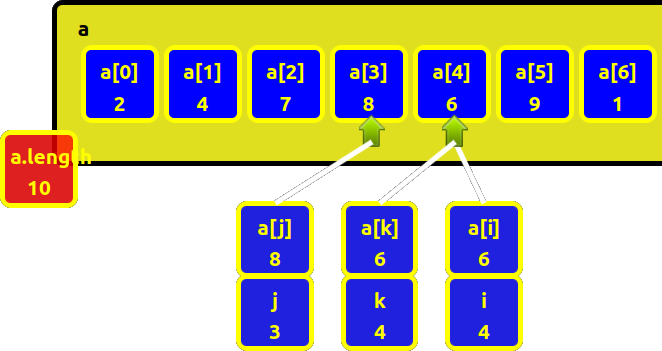}
\end{tcolorbox}
\caption{Indexes {\tt j} and {\tt k} introduced, and will identify adjacent cells to be swapped.}
\label{fig:affectation10}
\end{minipage}
\Description{AGT Insertion}
\end{figure}

The data in figure \ref{fig:insertion1} illustrates the general case, 
insertion within the sorted part. 
The array prefix is in order between indices 0 and 3, and the value to be inserted, 
6, is at index 4. 
The value 6 must be swapped first with the value 8, and then with the value 7, 
and finally placed between 4 and 7.
The algorithm  therefore works by iteration, exchanging the current value with the previous 
value until the 6 lands in its final place\footnote{We note the criticality of having chosen 
sample data that illustrates the need for iteration. Were the data insufficiently general, a 
novice might overlook the need for iteration altogether. Appropriate choice of data is thus a 
critical technique that must be taught, but that is assumed here.}.
Having recognized the need for iteration, we create a Loop macro whose sections 
we then develop in the order discussed above.

\paragraph{Loop body:}
We start by programming the loop body. The manipulations we do will, in effect, perform the first iteration of the loop, but must be done in a manner that generates correct code for the general case of an arbitrary step in the iteration. Were we to do the manipulation directly on the array elements themselves, this would not be the case because the code generated would be:
\begin{Verbatim}[samepage=true]
        Loop
            tmp = a[4] ;
            a[4] = a[3] ;
            a[3] = tmp ;
\end{Verbatim}
Accordingly, we must recognize the need to perform the swap indirectly using indexes, which we now create:
\begin{itemize}
\item[\textbullet] {\tt k}: the index of the value to insert, initialized to {\tt i}. The index
{\tt k} will progress to the left until landing at the final place of the value to be inserted;
\item[\textbullet] {\tt j}: the index of the value immediately 
to the left of {\tt a[k]}, that is, {\tt k-1}, which will be used to swap {\tt a[k]} 
with {\tt a[k-1]}\footnote{We introduce a separate index j because directly manipulating a[k-1] is not supported in AlgoTouch.}. 
\end{itemize}

We establish the appropriate initial values of {\tt k} and {\tt j}, as shown in 
Figure \ref{fig:affectation10}, either by opening dialog boxes for them and 
entering the values 3 and 4, respectively, or by directly manipulating k and 
j to have the correct values. 

\begin{figure}[htbp]
\centering
\begin{minipage}[b]{0.49\textwidth}
\centering
\begin{tcolorbox}[width=0.75\textwidth, colback=mygray, colframe=white, boxsep=0pt, left=2pt, right=-6pt, top=\topboxpadding, bottom=\bottomboxpadding]
\centering
\includegraphics[scale=0.25]{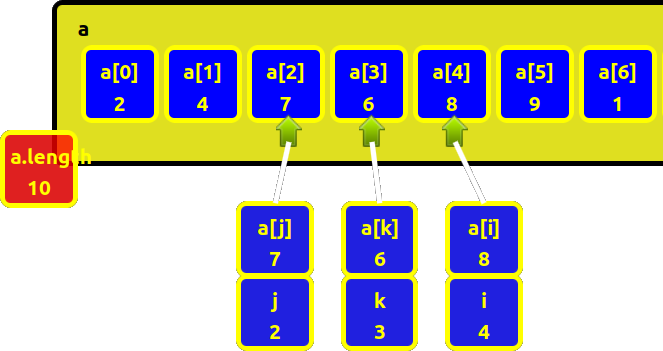}
\end{tcolorbox}
\caption{The values 8 and 6 have been swapped. The value 6 is still in the wrong place.}
\label{fig:insertion2}
\end{minipage}
\hfill
\begin{minipage}[b]{0.49\textwidth}
\centering
\begin{tcolorbox}[width=0.75\textwidth, colback=mygray, colframe=white, boxsep=0pt, left=1pt, right=-6pt, top=\topboxpadding, bottom=\bottomboxpadding]
\centering
\includegraphics[scale=0.25]{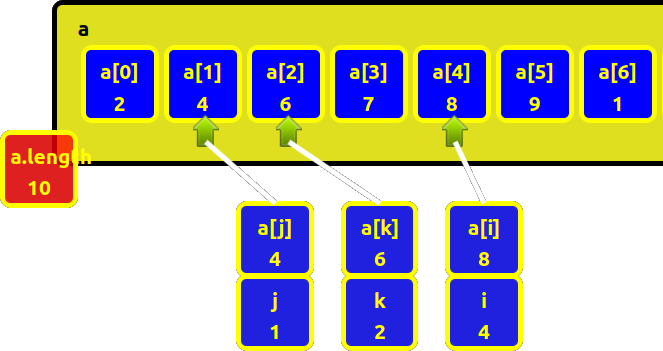}
\end{tcolorbox}
\caption{The values 7 and 6 have been swapped, and 6 is in its place.}
\label{fig:affectation11}
\end{minipage}
\vfill
\Description{AGT Insertion}
\end{figure}

Next, we select {\tt Loop} in the Instructions Area, and start Recording mode. 
Then, we perform the swap action, which moves the element being inserted left one place. 
By performing the swap indirectly via the indexes,  and then decrementing the indexes, 
we effect the manipulations required of the first iteration 
(as shown in Figure \ref{fig:affectation11}), and generate the code:
\begin{Verbatim}[samepage=true]
        Loop
            tmp = a[k] ;
            a[k] = a[j] ;
            a[j] = tmp ;
            j = j - 1 ;
            k = k - 1 ;
\end{Verbatim}
The concept of programming a generic iteration step with the direct manipulation of a specific step must be learned.

At this stage, as shown in Figure \ref{fig:insertion2}, the value 6 
is still not correctly placed, so it is necessary to repeat the previous operations.
We can do so, and simultaneously validate the generality of the loop body, by turning 
Recording mode off, and clicking the Execute in Construction mode button on each line of the 
Loop block, one after the other.
After completing the second iteration, the value 6 in {\tt a[k]} is correctly placed between 4 in {\tt a[j]} and 7 in {\tt a[3]}, as shown in Figure \ref{fig:affectation11}.

Now we need to enter code that will prevent the loop from further executing.
\paragraph{Loop exit:}
We define the exit condition by selecting {\tt Until}, turning Recording mode on,
comparing the two values {\tt a[j]} and {\tt a[k]}, 
and choosing the condition {\tt a[j]~<=~a[k]}. 
This generates the code of the {\tt Until} block:
\begin{Verbatim}[samepage=true]
        Until
            a[j] <= a[k]
\end{Verbatim}
We can validate the generated code by turning Recording mode off, and executing the {\tt Until} in Construction mode, which correctly advances the cursor to Terminate because the condition holds in the current program state.

\begin{figure}[htbp]
\centering
\begin{minipage}[t]{0.49\textwidth}
\centering
\begin{tcolorbox}[width=0.75\textwidth, colback=mygray, colframe=white, boxsep=0pt, left=2pt, right=-6pt, top=\topboxpadding, bottom=\bottomboxpadding]
\centering
\includegraphics[scale=0.25]{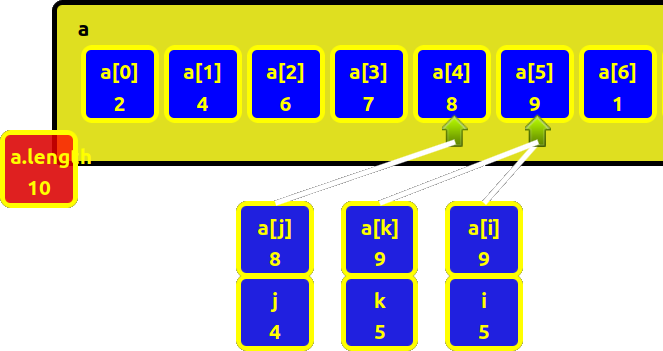}
\end{tcolorbox}
\caption{The value 9 is to be inserted into the sorted part of the array. It is already in the correct place.}
\label{fig:insertion3}
\end{minipage}
\hfill
\begin{minipage}[t]{0.49\textwidth}
\centering
\begin{tcolorbox}[width=0.75\textwidth, colback=mygray, colframe=white, boxsep=0pt, left=7pt, right=0pt, top=\topboxpadding, bottom=\bottomboxpadding]
\centering
\includegraphics[scale=0.25]{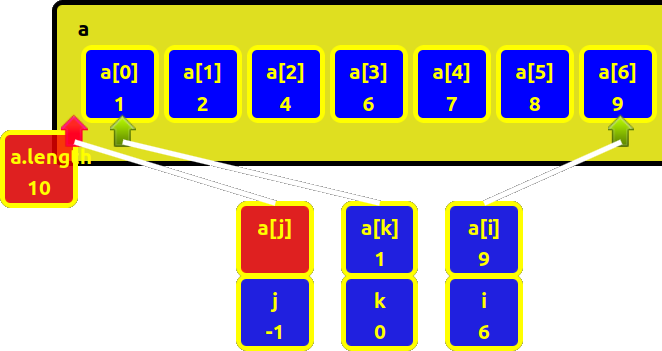}
\end{tcolorbox}
\caption{After several executions, in Construction mode, of the loop body, the value 1 is in the correct place.}
\label{fig:insertion4}
\end{minipage}
\vfill
\Description{AGT Insertion}
\end{figure}

\paragraph{Loop initialization:}
At this point, we code the loop initialization. 
To do so, we select {\tt From}, turn Recording mode on, copy {\tt i} to {\tt k}, and 
copy  {\tt i - 1} to {\tt j}. 
This sets {\tt j} and {\tt k} to the same initial values that we previously established manually with Recording mode off, and generates the initialization code:
\begin{Verbatim}[samepage=true]
        From
            k = i ;
            j = i - 1 ;
\end{Verbatim}

\paragraph{Handling of special cases:}
The program produced so far only handles insertion within the sorted part. 
We now consider the two special cases: 
(1) insertion at the end of the sorted part, and 
(2) insertion at the beginning of the array.
\begin{enumerate}
\item 
When the value in {\tt a[i]} is greater than or equal to {\tt a[i-1]}, it is already 
in its correct location from the get-go, there is nothing to do. 
This case is correctly handled by the general case because the exit condition 
{\tt a[j]~<=~a[k]} is immediately true, thereby avoiding execution of the Loop block. 
We can validate correct execution of the code in either of two ways: first, we can just execute the macro with i still 4, as shown in Figure \ref{fig:affectation11}. 
The initialization code in  {\tt From} block will reset {\tt j} and {\tt k} appropriately, and 
(because {\tt a[0..i]} has previously been ordered) the condition in  {\tt Until} block
will immediately skip the {\tt Loop} block. 
Alternatively (because the sample data was chosen to contain a 9 in {\tt a[5]}), 
we can change {\tt i} to 5, as shown in Figure \ref{fig:insertion3}, and execute the macro.

\item 
When the value of {\tt a[i]} is less than the value of {\tt a[0]}, it must be inserted at 
the beginning of the array. 
Because the sample data was chosen to have {\tt a[6]} be 1, this case can be validated 
by changing {\tt i} to 6. 
After executing, in Construction mode, the {\tt From} block, which initializes indexes {\tt j} and {\tt k}
appropriately, running the {\tt Loop} block five times swaps values until 
the 1 is positioned at {\tt a[0]}, as shown in Figure \ref{fig:insertion4}. 
An additional execution of the {\tt Loop} block would cause an error 
because {\tt j} is -1 and {\tt a[-1]} does not exist, 
which AlgoTouch indicates by displaying {\tt a[j]} in {\bf red}. 
It is therefore essential to add a second clause to the exit condition that stops the iteration.
To do so, in Recording mode, we select the {\tt Until} block, compare the value of {\tt j} to 0, 
and select the condition {\tt j~<~0}. 
Code for the additional exit condition is added to the {\tt Until} block:
\begin{Verbatim}[samepage=true]
        Until
            j < 0
            a[j] <= a[k]
\end{Verbatim}

To avoid subscript errors, AlgoTouch automatically arranges the order of exit conditions so that index values are tested first. 
Thus, as illustrated in the generated code, {\tt j~<~0} is tested before {\tt a[j]~<=~a[k]} 
not by virtue of careful cursor placement, but because of the ordering rules built 
into code generation.

Alternatively, we could have chosen {\tt k~<=~0} as the exit condition. 
However, following the principle of {\bf data-state visibility} articulated 
in section \ref{sec:objectives}, the display of {\tt a[j]} in red when {\tt j} is -1 
warns of the error case, and thus encourages testing {\tt j} rather than {\tt k}. 
Such visual highlighting simplifies data interpretation, and strengthens a user’s intuition when faced with sensitive boundary conditions.
\end{enumerate}

\paragraph{Handling multiple exit conditions:}
As illustrated in the previous example, exit conditions are not always placed in the order of their generation. 
AlgoTouch automatically arranges 
the order of conditions so that index values are tested first.
This functionality is possible because AlgoTouch knows how to identify 
the indexes and associated constraints.

\paragraph{Processing at loop exit:}
For the {\tt Terminate} block, there is nothing to code in this example because, in all cases, the array is already sorted.
\paragraph{Final program:}
The final program is as follows:
\begin{tcolorbox}[colframe=black, colback=white, boxrule=0.5pt]
\begin{Verbatim}[samepage=true]
        Define InsertElt
            From
                k = i ;
                j = i - 1 ;
            Until
                j < 0
                a[j] <= a[k]
            Loop
                tmp = a[k] ;
                a[k] = a[j] ;
                a[j] = tmp ;
                j = j - 1 ;
                k = k - 1 ;
            Terminate
        End
\end{Verbatim}
\end{tcolorbox}
\label{code:insert1}

The final code was produced in stages, without respecting the order of 
the Macro Loop blocks. 
The {\tt Loop} block was completed first, then the {\tt Until} block, 
and finally the {\tt From} block. 
This example illustrates the notion of {\bf non-linearity}, defined 
in the \ref{sec:objectives} section, specific to AlgoTouch, 
where each block is treated independently.

\subsection[Construction of programs with multiple macros  ]{Construction of programs with multiple macros } \label{sec:prog_complexes}


As presented so far, writing an AlgoTouch program consists of assembling basic 
elements—input/output, assignments, and conditionals—in a single Simple Macro or Loop macro.
The construction of programs with nested loops requires macro calls.
This section first describes how to call a macro, and illustrates how macros can be used in a bottom-up approach for developing nested loops.
It then discusses the use of direct manipulation to design programs top-down. 

\subsubsection[Macro call]{Macro call}
A defined macro can be executed as a new indivisible operation by clicking its black triangular button (illustrated in Figure \ref{fig:macro}).
If Recording mode is on, this generates a call instruction at the cursor’s location in the 
Instruction Area, and performs the data manipulations specified in the macro.
For example, if the macro {\tt MyMacro} is defined, a call to it within the macro {\tt Main} can be generated by executing {\tt MyMacro} in Recording mode. 
The resulting {\tt Main} macro code becomes then:
\begin{tcolorbox}[colframe=black, colback=white, boxrule=0.5pt]
\begin{Verbatim}[samepage=true]
//
// Example of macro call
//
Define Main
    Do
        MyMacro ;
    End
\end{Verbatim}
\end{tcolorbox}
\label{code:Principale}

Subsequent data manipulations performed, with Recording mode still on, would generate additional instructions after the call. 

\subsubsection{Bottom-up Approach}
This standard approach involves first generating code for a macro, and then calling it in another macro being created.

\paragraph{General case:}
For this example, we define a new macro {\tt DisplayAvg} that reads two integers, calculates their average by invoking the {\tt Average} macro (previously defined in the section \ref{sec:simpleMacro}), and displays the result. 
The code of the {\tt DisplayAvg} macro, after recording the actions defined above, is as follows:

\begin{tcolorbox}[colframe=black, colback=white, boxrule=0.5pt]
\begin{Verbatim}[samepage=true]
//
// Computes and displays the average of two integers entered on the keyboard
//
Define DisplayAvg
    Do
        Read "Value of a " a ;
        Read "Value of b " b ;
        Average ;
        Write "Average value : " avg ;
    End
\end{Verbatim}
\end{tcolorbox}
\label{code:displayAverage}

\paragraph{Nested loops:}
In most programming languages, nested loops can be written one inside another. In contrast, each loop is defined in AlgoTouch as its own Loop Macro, where the outer loop invokes the inner loop with a call instruction. An advantage of this decomposition is that each macro can be developed and tested individually in a bottom-up manner \cite{Kollig2013}. 

We illustrate the nesting of Loop Macros by implementing Insertion Sort on the assumption that its inner loop has already been implemented and thoroughly tested as macro {\tt InsertElt} (as developed in section \ref{sec:loop_construction}). Recall that {\tt InsertElt} inserts the value located at index {\tt i} into the already sorted sub-array {\tt a[0..i-1]}. 
To code the outer loop, we first manually set {\tt i} to 1\footnote{We might have set {\tt i} to 0 if we had overlooked that a sub-array of one cell is already sorted. This would cause no harm.}, and reset the values in the array {\tt a} to the order shown in Figure \ref{fig:insertion1} (with Recording mode off).

Then, we use direct data manipulation to repeatedly execute {\tt InsertElt} and increment {\tt i} until the last value in the array has been inserted. The first iteration is effected, and code is generated for the loop body, by performing {\tt InsertElt} and incrementing {\tt i} once (with the cursor positioned at the {\tt Loop} block, and Recording mode on). 

The remaining iterations are effected by turning Recording mode off, and repeatedly executing, in Construction mode, the generated code. The moment to stop iterating and define the exit condition is visually signaled when array index {\tt i} lights up in red because it has gone out of bounds. We then turn Recording mode on again, and insert the exit condition {\tt i >= a.length} at the {\tt Until} block. 

The final program produced is as follows: 

\begin{tcolorbox}[colframe=black, colback=white, boxrule=0.5pt]
\begin{Verbatim}[samepage=true]
        Define InsertionSort
            From
                i = 1 ;
            Until
                i >= a.length
            Loop
                InsertElt;
                i = i + 1 ;
            Terminate
        End
\end{Verbatim}
\end{tcolorbox}
\label{code:triInsert}

\subsubsection{Top-down approach: by simulating called macros}
\label{sec:top-down}
The process combines code execution with user-driven data manipulation. 
The user may insert calls to macros whose internal code and
algorithm are not yet defined and will be developed later. 
During the execution of the main macro, AlgoTouch pauses whenever
it encounters an empty macro\footnote{A created macro whose code is not defined.}. 
At that point, control is handed over to the user, who must 
manually produce the data manipulations corresponding to the 
expected result of that macro. 
Once this has been done, the user resumes execution, which 
proceeds until the next undefined macro is reached.
In the example of the previous section, suppose that the {\tt InsertElt} macro has not been coded: each time it is called, the user must manually insert the value into the correct position in the array.

The advantage of this approach is that it allows the user to verify that the logic of their program is correct while freeing them from the constraint of having to think about how to generate each macro.

This operation differs from the ``{\em stubs}'' approach proposed 
in \cite{caspersen2009stream} where the call to a function not yet developed
produces a predefined result, identical to each call. 
With AlgoTouch, it can be different for each call depending 
on the manipulations performed. 
The user will then have to generate the macro instructions. 
This original way of proceeding allows a complex program to be built in successive stages as detailed in \cite{michel2021construction} and provides the user with greater flexibility in design.

\subsection[Execution and Visualization]{Execution and Visualization} \label{sec:execution}
In the previous sections, we have frequently discussed the notion of 
execution, particularly in relation to generation of code. 
Building structures such as conditionals or loops does indeed involve 
re-executing partial code to complete missing elements, 
such as an {\tt else} block, an {\tt Until}, or a {\tt Terminate} block. 
During the construction phase, the processes of data manipulation, code generation, 
and execution are closely interdependent.

Once the construction phase is complete, the user should also be able to execute 
their program to verify its proper functioning. 
AlgoTouch offers several execution modes:
\begin{itemize}
\item[\textbullet] Construction mode \
\item[\textbullet] Direct mode \
\item[\textbullet] Animation mode
\item[\textbullet] Detailed or non-detailed mode \
\end{itemize}
The first two modes are essential for generating programs 
by manipulating data. 
Construction mode is a unique mode specific to AlgoTouch, 
which allows execution of code segments. 
Direct mode allows execution of all the code in a macro.
The last two modes were added to provide automatic viewing options for
the complete execution of the program.

\subsubsection[Execution in Construction mode]{Execution in Construction mode}
This mode, already mentioned in the previous examples, is used during program creation.
It allows the user to execute a specific instruction or block
({\tt Do}, {\tt From}, {\tt Until}, {\tt Loop} or {\tt Terminate}). 
This mode offers several possibilities:
\paragraph{Complete the conditionals:} 
If, during execution, AlgoTouch encounters an empty
{\tt else} block, it suspends execution and prompts the user 
to complete the code. 
The goal is to progressively complete all undefined parts until the code is fully specified.
This works similarly to Pygmalion \cite{smith1975pygmalion} and ManipoSynth
\cite{hempel2022maniposynth}, as mentioned in section \ref{sec:conditional}.

\paragraph {Execute loop bodies until an exit condition:} 
Exit conditions can be reached when the solution to the problem is found 
or when a new execution of the loop body causes an error.
AlgoTouch allows the user to execute the body of the loop, 
checking the values of the variables, until one of these results is obtained
(section \ref{sec:method}).

\paragraph{Use reactivity:} 
As the code is generated, the user can view the effects of the execution 
of each instruction on the data. 
This reactivity makes it possible to validate the fact that 
the created instructions produce the expected results. 
Simply select a line of code and execute it in
Construction mode.

\paragraph {Debug:} 
The user can execute the program step by step from the beginning to more easily spot errors. 
It is also possible to execute the program from a specific instruction 
after defining the data content. 
Unlike a classic debugger, there is no need to set a breakpoint 
and start the full execution. 
This unique approach makes it easier to test section of code.

\paragraph{Top-down approach:} 

It is through execution in Construction mode that AlgoTouch implements the Top-Down approach described in section \ref{sec:top-down}.
The execution stops at the call of the empty macro, 
giving the user the possibility of simulating the actions to be performed. 

\subsubsection[Direct Mode Execution]{Direct Mode Execution}
Direct execution allows a macro to be executed in its entirety, 
by clicking directly on the black triangle of
the macro icon on the Workspace. 
This is the usual execution mode for a program.

\subsubsection[Other execution modes]{Other execution modes}
The main research objective of AlgoTouch is to show that it is possible to build 
programs by directly manipulating data. 
However, from the user's point of view, it is important to be able 
to easily visualize program execution. 
Taking advantage of the graphical interface, in particular the built-in
visualization of data, and the execution engine, it was easy 
to add additional execution modes.

\paragraph{Execution in animation mode:}
Animation mode is used to visualize the evolution of data in 
the Workspace Area during the execution of a macro.
The user can choose block-by-block execution ({\tt Do}, {\tt From}, 
{\tt Until}, {\tt Loop}, {\tt Terminate}) or instruction by instruction.
Execution is performed in step-by-step mode or in automatic mode.
In automatic mode, blocks or instructions are executed at regular intervals.
The current block or instruction is highlighted, and the display of the data
contents is updated as it progresses. 
A slider allows the user to adjust the speed execution mode. 
This mode is particularly useful for tracking data changes during 
macro execution. 
It is valuable in the context of demonstrating the functioning of algorithms.

\paragraph{Execution in detailed or non-detailed mode:}
All executions can be performed in detailed or non-detailed mode. 
For example, in direct execution mode, although the program runs completely,
the detailed mode shows the evolution of the data content of the
program.
Conversely, in non-detailed mode, the state of the variables 
is only displayed at the beginning and end of execution. 
By default, AlgoTouch operates in detailed mode.

\section[Evolution]{Evolution}
The initial version of AlgoTouch dates back to 2014. 
Over the years, AlgoTouch has undergone continuous development and 
refinement, guided by practical experience and feedback from users,
including students, teachers, and researchers. 
This evolution encompasses both technical improvements and usability enhancements, aiming to make the tool more intuitive 
while preserving its core principles.

\subsection[Technical developments]{Technical developments}
The first version of AlgoTouch was developed in Java. 
Using it required a Java virtual machine and downloading a ``jar'' file. 
For many users, the mere download requirement was a barrier. 
In addition, each update required a new download. 
These technical limitations were a major obstacle to using AlgoTouch. 

To facilitate accessibility, we subsequently opted for a web-based solution. 
Users can now instantly access the latest version of AlgoTouch. 
In addition, users have the ability to launch AlgoTouch directly on a particular program by passing 
parameters in a URL in the standard JSON format. 
This mechanism, inspired by Python Tutor \cite{guo2013online}, facilitates the distribution 
of demonstration programs. 

The current version is designed in ReactJS, which has made it possible to benefit from the numerous available libraries, and to accelerate the development and distribution of the new version. 

\subsection[Editing Features]{Editing Features}
From our experience, and in agreement with Hempel \cite{hempel2022maniposynth},
we found it important to provide explicit access to the generated code with the ability to modify it.
Indeed, even when it is possible to achieve desired effects by direct manipulation, code is an unambiguous way to record the actions performed. 
It allows the user to inspect the logic of past operations, understand them, and make desired changes. 

\subsubsection[Line-by-Line Code Editing]{Line-by-Line Code Editing}
In early versions of AlgoTouch, the emphasis was on code generation by direct manipulation 
followed by execution of the program. 
When an operation proved to be unsatisfactory, the only editing possibility consisted of erasing 
an entire block ({\tt Do}, {\tt From}, {\tt Loop}, {\tt Until}, {\tt Terminate}) and re-manipulating 
the data to generate new code. 
This design was a major inconvenience, as it was necessary either to generate the code without 
errors or to start all over again. 

Accordingly, we now allow an instruction line in the code to be selected, and then permit the user to add instructions after the line by direct manipulation, or execute the instruction, or delete it. 

\subsubsection[Simplification of Conditional Generation]{Simplification of Conditional Generation}

Our experiments and surveys showed that our original mechanism for building a conditional was sometimes
overly restrictive for users. 
Now, to complete the {\tt else} clause of a conditional, there are two choices: 
use the execution method presented in the section \ref{sec:conditional} , or directly 
select the keyword {\tt else} and add code by manipulation. 
Furthermore, it is now possible to delete or add an {\tt else} clause, as necessary. 

Recall from section 3.6 that the test of an {\tt if} statement is generated on the assumption that for the current data, the true branch is to be executed. But if operations are only required when the generated condition is false, this mechanism leads to an unsightly empty true branch. To deal with this situation, AlgoTouch offers a refactoring command to improve code readability by inverting the condition.
For example, the command would restructure this code fragment 
\begin{Verbatim}[samepage=true]
        if (v >= 0) { 
            // v >= 0 
        } else { // v < 0 
            v = 0; 
        }
\end{Verbatim}
as follows:  
\begin{Verbatim}[samepage=true]
        if (v < 0) {
            v = 0 ;
        }
\end{Verbatim} 

\subsubsection[Bimodality]{Bimodality}
These code editing features can be compared to the bimodality 
presented by Hempel in his thesis \cite{hempel2022maniposynth}: 
the user can work directly on the data, or on the code in text form.
While AlgoTouch is based on the idea of manipulating data based on its current values to achieve 
a desired objective ({\bf value-centered} principle)), these editing facilities deviate 
from that principle—a necessary and beneficial compromise that improves the flexibility 
of code production. 
However, we retain the core principle that instructions can only be added by manipulating the data. 

\subsection[Use of ``mainstream'' programming languages]{Use of ``mainstream'' programming languages}
\label{sec:languages}
To implement AlgoTouch, we designed a simple yet Turing-complete programming language called AGT. 
Although minimal, this new language may be an obstacle for users who prefer to use a standard language
like C, C++, Java, or Python. 
To avoid alienating such users, any of those programming languages can be selected to replace AGT.
This flexibility emphasizes that a programming language is only one of many means for expressing an algorithm, which is independent of the specific syntax of code. 

\begin{figure}[htbp]
\centering
\begin{tcolorbox}[
colback=white, 
colframe=black, 
arc=4pt, 
boxrule=.5pt, 
boxsep=5pt, 
left=2pt, right=2pt, top=2pt, bottom=2pt, 
width=\textwidth 
]
\begin{minipage}[b]{0.49\textwidth}
\begin{tabular}{m{0.9\textwidth}}
\textbf{AGT Program}\\\hline
\begin{Verbatim}
Define InsertElt
    From
        k = i ;
        j = i - 1 ;
    Until
        j < 0
        a[j] <= a[k]
        
        
    Loop
        tmp = a[k] ;
        a[k] = a[j] ;
        a[j] = tmp ;
        j = j - 1 ;
        k = k - 1 ;
    Terminate
End
\end{Verbatim}
\end{tabular}
\end{minipage}
\hfill
\begin{minipage}[b]{0.49\textwidth}
\begin{tabular}{m{0.9\textwidth}}
\textbf{Equivalent in Python}\\\hline
\begin{Verbatim}
# Macro InsertElt
# Initialization
k = i
j = i - 1
# Exit conditions:
# Exit if j < 0
# Exit if a[j] <= a[k]
#
while (j >= 0 and a[j] > a[k]):
    # Loop body
    tmp = a[k]
    a[k] = a[j]
    a[j] = tmp
    j = j - 1
    k = k - 1
# Termination
#
\end{Verbatim}
\end{tabular}
\end{minipage}
\end{tcolorbox}
\begin {comment}
\begin{tabular}{|m{.47\textwidth}|m{.4775\textwidth}|}

\hline
\textbf{AGT Program} &
\textbf{Equivalent in Python}\\hline
\begin{Verbatim}
Define InsertElt
    From
        k = i ;
        j = i - 1 ;
    Until
        j < 0
        a[j] <= a[k]
 
    
    Loop
        tmp = a[k] ;
        a[k] = a[j] ;
        a[j] = tmp ;
        j = j - 1 ;
        k = k - 1 ;
    Terminate
End
\end{Verbatim}
&
\begin{Verbatim}
# Macro InsertElt
# Initialization
k = i
j = i - 1
# Exit conditions:
# Exit if j < 0
# Exit if t[j] <= t[k]
#
while (j >= 0 and t[j] > t[k]):
    # Loop body
    tmp = t[k]
    t[k] = t[j]
    t[j] = tmp
    j = j - 1
    k = k - 1
# Termination
#
\end{Verbatim}

\\\hline
\end{tabular}
\end{comment}
\caption{Instrumentation of the Python code of the macro {\tt InsertElt}.
The {\tt From}, {\tt Until}, {\tt Loop}, and {\tt Terminate} blocks are replaced by selectable comments. The {\tt Until} loop is replaced by
a {\tt while} loop.}
\label{tab:AGT_Python}
\Description{AGT Python}
\end{figure}
The {\tt From},
{\tt Until}, {\tt Loop}, {\tt Terminate} blocks of AGT do not exist in other languages, so AlgoTouch 
identifies them using selectable comments, i.e., {\tt \#~Initialization}, {\tt \#~Exit conditions},
{\tt \#~Loop body}, and {\tt \#~Termination}, as shown in Figure \ref{tab:AGT_Python}. 
Similarly, the exit conditions of an {\tt Until} block are associated with individual selectable
comments that are distinct from the continuation condition of a while statement formed as the
conjunction of the negated exit conditions.  
For example, the exit conditions {\tt j~<~0} and {\tt a[j]~<=~a[k]} in Figure \ref{tab:AGT_Python} 
are displayed as separate comments, and are translated to the while-loop continuation condition 
{\tt j~>=~0~and~a[j]~>~a[k]}. 
When an exit condition is selected, AlgoTouch highlights the associated continuation condition, 
and vice versa. 
This dual visualization allows for a better understanding of the correspondence between 
exit conditions and continuation conditions. 

As described in section \ref{sec:prog_complexes}, each loop is defined in AlgoTouch as a separate Loop
Macro that, regardless of programming language, is displayed individually in the Instructions Area. 
For languages other than AGT, when code is exported or echoed to the Console, macros are 
inline expanded, and comments that identify blocks and exit conditions are removed 
so the code conforms to the usual representation. 
For example, the AGT code for insertion sort developed above would be exported as the following Python code: 

\begin{tcolorbox}[colframe=black, colback=white, boxrule=0.5pt]
\begin{Verbatim}[samepage=true]
#
# Performs insertion sort
#
i = 1
while (i < len(a)) :
    # -> InsertElt: Inserts a[i] into the sorted part of a from 0 to i-1
    k = i
    j = i - 1
    while (j >= 0 and a[j] > a[k]):
        tmp = a[k]
        a[k] = a[j]
        a[j] = tmp
        j = j - 1
        k = k - 1
    # <- InsertElt
    i = i + 1
\end{Verbatim}
\end{tcolorbox}
\section[Assessment ]{Assessment}

This section presents examples of programs developed using AlgoTouch, educational experiments, and teacher surveys, illustrating its usability and pedagogical value.

\subsection[Examples of developed programs]{Examples of developed programs}
\label{sec:liste_prog}
To validate the approach, numerous programs were created with AlgoTouch. 
They correspond to examples and exercises from introductory courses on algorithms and programming. 
The developed examples are listed in Table \ref{tab:prog_list}.

\begin{table}
\centering
\caption{Programs developed with AlgoTouch. The programs are classified by category (conditions, iterations, use of arrays, sorting, etc.).}
\label{tab:prog_list}

\begin{tabular}{m{2cm}m{3.1cm}m{10cm}}
\hline
\textbf{Conditions} &
Parity &
Tests the parity of an integer. \\
& Leap year &
Determines if a year is a leap year. \\
& Next second &
Add one second to the time. \\
& Ticket price &
Calculates a ticket price based on the customer's age. \\
\textbf{Iterations} &
Grade control&
Controls the entry of a grade between 0 and 20. \\
& Factorial &
Calculate the factorial. \\
& Sequence Average &
Calculates the average of a sequence of values ending in -1. \\
& GCD &
Calculates the GCD of two integers. \\
& Syracuse &
Calculate the Syracuse sequence. \\
& Mystery number &
Guess a number between two limits. \\
& Timing &
Calculates the speed and distance covered by a runner. \\
\textbf{Arrays} &
Right shift &
Shifts values in an array one position to the right. \\
& Circular rotation &
Performs circular rotation of the values in an array. \\
& Insert element &
Inserts a[i] into the sorted part of array a up to i-1. \\
& Place the max &
Place the largest value from i into a[i]. \\
& Even and odd  &
Stores an array with alternating even and odd values. \\
& Unique &
Replaces unique values at the beginning of the array. \\
& Identical &
Tests if the values in the array are all the same. \\
& Anagram &
Checks if two character arrays are anagrams of one another.\\
& Hanged man &
Guess a word by suggesting a sequence of letters. \\
& Compress &
Replaces multiple values with value and occurrence. \\
\textbf{Binary} &
Integer to binary &
Converts an integer to an array of binary values. \\
& Binary to Integer &
Converts an array of binary values to an integer. \\
& Binary add &
Adds two arrays of binary values in binary. \\
& Binary project &
Calculates in binary the sum of 2 entered integers. \\
\textbf{Traversal} &
Minimum index &
Determines the index of the smallest element in an array. \\
& Minimum search &
Determines the value and index of the smallest element. \\
& Rain fall &
Calculates the average precipitation heights. \\
& Inc and dec &
Increment values > 0, decrement values < 0. \\
& Week temperature &
Calculates the average temperature over a week. \\
& Interlacing &
Checks for alternating even and odd values. \\
\textbf{Searching} &
Sequential Search &
Sequential Search in an array. \\
& Sentinel Search &
Search with forced sentinel as last element. \\
& Binary Search &
Binary Search in a sorted array. \\
\textbf{Sorting} &
Bubble Sort &
- \\
& Insertion Sort &
- \\
& Selection Sort &
- \\
& Heap Sort &
- \\
& Merge Sort &
Merging two sorted arrays. \\
& Partition Sort &
Non-recursive QuickSort. \\\hline
\end{tabular}
\normalsize
\end{table}
This table indicates the diversity of exercises that are easily expressed in AlgoTouch 
despite its limited functionality.
A wide range of programming skills, from basics to more advanced concepts like sorting and searching, 
are covered.

\subsection[Educational experiments]{Educational experiments}
We conducted a number of experiments with students and teachers. 
Considering our environment and the constant evolution of the tool, certain experiments described 
in this section do not claim to meet all the expected scientific criteria \cite{ko2015practical}. 
Their primary objective was to obtain rapid feedback from users and better develop improved 
features of AlgoTouch. 

\subsubsection[Experiments with students]{Experiments with students}
A first experiment \cite{adam2019direct} was conducted with junior students from the Computer Science
Department of the IUT of Vannes, University of South Brittany, to compare the standard approach 
to writing code with direct manipulation programming. 
The experiment was conducted with a sample of 54 junior students divided into 2 groups. 
The first group programmed with Python Tutor \cite{guo2013online}, 
an online tool for entering a program and visualizing its execution, 
and the second group programmed with AlgoTouch. 

For a first category of exercises, without iteration, requiring only a sequence of calculation 
instructions, the results are similar for both tools. 
A second category of exercises involved iterations, and included well-known calculations 
(GCD by the subtractive method, power) and exercises of increasing difficulty 
for programming algorithms on arrays. 
The results demonstrated higher success and validity rates for programs created with 
the AlgoTouch tool than with the Python Tutor tool.

Other informal experiments were conducted with different subjects. 
The authors of the paper used AlgoTouch as part of their 
undergraduate teaching. 
The students' feedback motivated the improvements mentioned above.

\subsubsection[Experiments with teachers]{Experiments with teachers}
A survey was conducted in 2017 among middle school teachers (70) who had participated in AlgoTouch workshops in Brittany. 
The question asked was: ``What are, {\em a priori}, the strong points 
of the AlgoTouch software''? 
The answers, in the form of keywords, ranked in order of importance, are:
\begin{enumerate}
\item Visual - Direct visualization (in real time)
\item Direct manipulation - Simple manipulation
\item Transition to a ``real'' programming language - 
      Link with unplugged activity.
\item Automatic code generation
\item Machine Operation - Normal Mode vs. Blind Mode
\item Good approach to algorithms - Simple approach - Gentle introduction to the language
\item Better understanding of the principle and operation of algorithms
\end{enumerate}

\subsubsection[Experience of a university teacher]{Experience of a university teacher}
A teacher, not participating in our research, used AlgoTouch 
in a course called ``Algorithms and Imperative Programming''\footnote{Similar 
to the CS1 course.} for first-year undergraduate students. 
She shared the following comments with us:
\begin{itemize}
\item[\textbullet] AlgoTouch clearly illustrates fundamental programming 
concepts such as loops, arrays, and the use of subroutines via ``macros'';

\item[\textbullet] Using AlgoTouch to teach these principles undoubtedly 
reinforces understanding for students, thus facilitating the learning 
of a specific language such as the C language;

\item[\textbullet] A support tool for those who are new to programming 
as well as for those with prior knowledge in algorithms which gives them 
the opportunity to learn other languages via the Export Areas;

\item[\textbullet] We clearly observe the positive developments between 
the two versions, Java and JavaScript, filling in the gaps in the Java 
version\footnote{The first version of AlgoTouch was written in Java, 
the second is a web version written in JavaScript with React.}. 
For example, the ability to edit a part of the code is a nice improvement, 
as is the GUI~;

\item[\textbullet] However, fully mastering the tool requires a special effort from both the instructor and the learners. I believe it's crucial to dedicate additional hours, outside of the curriculum, to support struggling students in achieving their goals. This point illustrates the difficulty of adopting a new approach to programming.
\end{itemize}

\section{Conclusion and perspectives}
This article presented an original tool for building programs by direct data manipulation. Unlike the traditional approach of entering lines of code into a programming language editor and proceeding by successive executions, AlgoTouch adopts a visual approach to building programs by direct-data-manipulation operations that correspond to the basic operation of a computer: assignments, arithmetic operations, conditions, repetitions of a set of operations. To allow the user to gradually learn a programming language, each operation generates instructions in a programming language (AGT, Python, C, C++, or Java). This approach avoids syntax and semantics errors. The user can ``play'' with the data and envision an algorithm to solve the problem. Because the system is visual, the user always sees what they are doing, and can correct or redo operations to achieve the given objective. 

Once the algorithmic solution is found, the user may want to keep it in the form of an executable program. To this end, AlgoTouch records data manipulation as code, and thus gradually builds the program. One of the original features of AlgoTouch is to process conditionals and loops only through direct data manipulation. Furthermore, code modification is feasible, but still in data-manipulation mode.

The visual nature of the system is not limited to data manipulation. 
Indeed, once a program is built, the user executes and visualizes 
its progress in the same Workspace. 
Since users have permanent access to the data and the program code, they can set the
values of certain data and launch an execution from any instruction 
in the code. 
They have the choice of executing a single instruction or a block of
instructions such as a loop body or a macro. 
These different possibilities allow great flexibility in testing.

Programs illustrating the concepts of conditionals, iterations, and array processing
have been developed. 
Most of the examples covered in this paper are problems encountered in courses
introductory programming courses (CS1).

Several experiments have shown the benefit of AlgoTouch for both students and teachers. For a beginner, as Bret Victor has observed \cite{victor2012learnable}, writing a program requires mentally imagining the result of each instruction. With AlgoTouch, in contrast, the user performs an operation, and sees both its result and the instruction generated. For a teacher, each programming concept (variable, assignment, operation, condition, iteration) can be fully visualized within the AlgoTouch system. In addition, to explain the principle of an algorithm, the teacher can visualize the execution of the corresponding program before coding it. 

The objectives defined in section \ref{sec:objectives} have been achieved:
\begin{enumerate}
\item \textbf{Value-centered}: 
\textit{the code is generated solely by data manipulation.}
All code is created through direct manipulation. 
However, the user can make changes in the program. 
While deleting code is supported, new instructions
are inserted only by data manipulation.

\item \textbf{Non-linearity}: 
\textit{the code can be generated partially depending on the values 
of the variables}. 
An AlgoTouch program can be executed even if pieces of code are missing 
(branches of conditionals, bodies of loops, bodies of called macros).

\item 
\textbf{Visibility of data and program state}: 
\textit{users must always have a visual representation of program data}. 
The program state is visible at each stage of execution. 
In particular, the different execution modes described in \ref{sec:execution} allow data to be viewed during program execution.

\item \textbf{Clear semantics of operations}: 
\textit{the system must allow for intuitive understanding of how the operations work}.
The operations performed by the user directly correspond to basic machine instructions. The interface can display the progress of each arithmetic operation, comparison, and access to an element of an array.

\item \textbf{Reactivity}: 
\textit{executing instructions modify the program data}.
AlgoTouch's various execution modes make it a sophisticated debugger. 
In addition to line-by-line or block-by-block program execution from the start of the program, it is also possible to resume execution at any line of program, whereupon the current values of the data are then used. 

\item \textbf{Turing-completeness}: 
\textit{AlgoTouch can generate Turing-complete code}.
Indeed, AlgoTouch’s use of variables and generation of conditionals and loops is sufficient for the expression of arbitrary computations. 
\end{enumerate}

In its current version, AlgoTouch should be considered as a proof of concept. Many enriched functionalities remain to be explored, particularly at the level of data and program structures. 

The data structures used by AlgoTouch are limited to variables and one-dimensional arrays, which are sufficient to illustrate the main algorithms presented in introductory programming courses. Other structures could be added like two-dimensional arrays, strings, lists, trees, stacks, and queues. The notion of type could also be extended to reals and Booleans. 

AlgoTouch programs are structured exclusively through macros with global variables.
While this approach allows writing a wide variety of algorithms, 
adding parameters to macros and managing variable scope  
would be a significant step forward. 
These improvements would allow for the integration of concepts similar to traditional languages, introducing the use of functions, procedures, and libraries, thus enriching AlgoTouch and making the tool more versatile for experienced users. 

\section*{Acknowledgments}
We would like to express our sincere gratitude to 
E. Sassi (Université de Bretagne Sud), 
E. Desmontils and M. Barraud (Université de Nantes), 
as well as C. Declercq and S. Hoarau (Université de la Réunion) 
for their valuable contribution using AlgoTouch in their teaching. 
We also thank M. McGuffin and C. Fuhrman (École de Technologie Supérieure,
ÉTS, Montréal) for the insightful discussions regarding visual programming 
and data manipulation. 
Special thanks are due to T. Teitelbaum (Professor Emeritus, Cornell
University) for his insightful advice on AlgoTouch, and for his careful 
review and correction of the English version of this article.
\bibliographystyle{plain}
\bibliography{biblio}

\begin{thebibliography}{10}

\bibitem{adam2019direct}
Michel Adam, Moncef Daoud, and Patrice Frison.
\newblock Direct manipulation versus text-based programming: An experiment
  report.
\newblock In {\em Proceedings of the 2019 ACM Conference on Innovation and
  Technology in Computer Science Education}, pages 353--359, 2019.

\bibitem{michel2021construction}
Michel Adam, Moncef Daoud, and Patrice Frison.
\newblock Construction d'un programme combinant ex{\'e}cution partielle et
  manipulation directe.
\newblock In {\em Atelier {\guillemotleft}Apprendre la Pens{\'e}e Informatique
  de la Maternelle {\`a} l'Universit{\'e}{\guillemotright}, dans le cadre de la
  conf{\'e}rence Environnements Informatiques pour l'Apprentissage Humain
  (EIAH)}, pages 25--33, 2021.

\bibitem{berry2013enseignement}
G~Berry, G~Dowek, S~Abiteboul, JP~Archambault, C~Balagu{\'e}, GL~Baron, C~de~la
  Higuera, M~Nivat, F~Tort, and T~Vi{\'e}ville.
\newblock L’enseignement de l’informatique en france. il est urgent de ne
  plus attendre.
\newblock {\em Rapport de l’Acad{\'e}mie des sciences}, page~1, 2013.

\bibitem{caspersen2009stream}
Michael~E Caspersen and Michael Kolling.
\newblock Stream: A first programming process.
\newblock {\em ACM Transactions on Computing Education (TOCE)}, 9(1):1--29,
  2009.

\bibitem{cheah2020factors}
Chin~Soon Cheah.
\newblock Factors contributing to the difficulties in teaching and learning of
  computer programming: A literature review.
\newblock {\em Contemporary Educational Technology}, 12(2):ep272, 2020.

\bibitem{chugh2016programmatic}
Ravi Chugh, Brian Hempel, Mitchell Spradlin, and Jacob Albers.
\newblock Programmatic and direct manipulation, together at last.
\newblock {\em ACM SIGPLAN Notices}, 51(6):341--354, 2016.

\bibitem{cooper2000alice}
Stephen Cooper, Wanda Dann, and Randy Pausch.
\newblock Alice: a 3-d tool for introductory programming concepts.
\newblock {\em Journal of computing sciences in colleges}, 15(5):107--116,
  2000.

\bibitem{cypher1993watch}
Allen Cypher and Daniel~Conrad Halbert.
\newblock {\em Watch what I do: programming by demonstration}.
\newblock MIT press, 1993.

\bibitem{fincher2020notional}
Sally Fincher, Johan Jeuring, Craig~S. Miller, Peter Donaldson, Benedict
  du~Boulay, Matthias Hauswirth, Arto Hellas, Felienne Hermans, Colleen Lewis,
  Andreas M\"{u}hling, Janice~L. Pearce, and Andrew Petersen.
\newblock Notional machines in computing education: The education of attention.
\newblock In {\em Proceedings of the Working Group Reports on Innovation and
  Technology in Computer Science Education}, ITiCSE-WGR '20, page 21–50, New
  York, NY, USA, 2020. Association for Computing Machinery.

\bibitem{frank2021low}
Ulrich Frank, Pierre Maier, and Alexander Bock.
\newblock Low code platforms: Promises, concepts and prospects. a comparative
  study of ten systems.
\newblock Technical report, ICB-Research Report, 2021.

\bibitem{fraser2015ten}
Neil Fraser.
\newblock Ten things we've learned from blockly.
\newblock In {\em 2015 IEEE blocks and beyond workshop (blocks and beyond)},
  pages 49--50. IEEE, 2015.

\bibitem{frison-hal-01753119}
Patrice Frison, Moncef Daoud, and Michel Adam.
\newblock {Transition didactique de l'activit{\'e} d{\'e}branch{\'e}e {\`a} la
  programmation avec AlgoTouch}.
\newblock In {\em {Didapro 7 -- DidaSTIC. De 0 {\`a} 1 ou l'heure de
  l'informatique {\`a} l'{\'e}cole}}, pages 1--17, Lausanne, Switzerland,
  February 2018.

\bibitem{graf2024assessing}
Oliver Graf, Sverrir Thorgeirsson, and Zhendong Su.
\newblock Assessing live programming for program comprehension.
\newblock In {\em Proceedings of the 2024 on Innovation and Technology in
  Computer Science Education V. 1}, ITiCSE 2024, page 520–526, New York, NY,
  USA, 2024. Association for Computing Machinery.

\bibitem{Kollig2013}
Channel Greenfoot.
\newblock The joy of code nb31: More loopiness.
\newblock \url{https://https://www.youtube.com/watch?v=ZdHtIigMbjA&t=273s},
  2013.

\bibitem{gulwani2016programming}
Sumit Gulwani.
\newblock Programming by examples: Applications, algorithms, and ambiguity
  resolution.
\newblock In {\em Automated Reasoning: 8th International Joint Conference,
  IJCAR 2016, Coimbra, Portugal, June 27--July 2, 2016, Proceedings 8}, pages
  9--14. Springer, 2016.

\bibitem{guo2013online}
Philip~J Guo.
\newblock Online python tutor: embeddable web-based program visualization for
  cs education.
\newblock In {\em Proceeding of the 44th ACM technical symposium on Computer
  science education}, pages 579--584, 2013.

\bibitem{halbert1984programming}
Daniel~Conrad Halbert.
\newblock {\em Programming by example}.
\newblock University of California, Berkeley, 1984.

\bibitem{hempel2022maniposynth}
Brian Hempel and Ravi Chugh.
\newblock Maniposynth: Bimodal tangible functional programming.
\newblock {\em arXiv preprint arXiv:2206.14992}, 2022.

\bibitem{hempel2019sketch}
Brian Hempel, Justin Lubin, and Ravi Chugh.
\newblock Sketch-n-sketch: Output-directed programming for svg.
\newblock In {\em Proceedings of the 32nd Annual ACM Symposium on User
  Interface Software and Technology}, pages 281--292, 2019.

\bibitem{hundhausen2002meta}
Christopher~D Hundhausen, Sarah~A Douglas, and John~T Stasko.
\newblock A meta-study of algorithm visualization effectiveness.
\newblock {\em Journal of Visual Languages \& Computing}, 13(3):259--290, 2002.

\bibitem{hundhausen2009can}
Christopher~D Hundhausen, Sean~F Farley, and Jonathan~L Brown.
\newblock Can direct manipulation lower the barriers to computer programming
  and promote transfer of training? an experimental study.
\newblock {\em ACM Transactions on Computer-Human Interaction (TOCHI)},
  16(3):1--40, 2009.

\bibitem{javier2021understanding}
Billy Javier.
\newblock Understanding their voices from within: difficulties and code
  comprehension of life-long novice programmers.
\newblock {\em International Journal of Arts, Sciences and Education},
  1(1):53--73, 2021.

\bibitem{johnson2023computational}
Chris Johnson.
\newblock Computational making with twoville.
\newblock {\em Journal of Computing Sciences in Colleges}, 38(8):39--53, 2023.

\bibitem{kadar2021study}
Rozita Kadar, Naemah~Abdul Wahab, Jamal Othman, Maisurah Shamsuddin, and
  Siti~Balqis Mahlan.
\newblock A study of difficulties in teaching and learning programming: a
  systematic literature review.
\newblock {\em International Journal of Academic Research in Progressive
  Education and Development}, 10(3):591--605, 2021.

\bibitem{ko2015practical}
Amy~J Ko, Thomas~D LaToza, and Margaret~M Burnett.
\newblock A practical guide to controlled experiments of software engineering
  tools with human participants.
\newblock {\em Empirical Software Engineering}, 20:110--141, 2015.

\bibitem{luxton2018introductory}
Andrew Luxton-Reilly, Simon, Ibrahim Albluwi, Brett~A Becker, Michail
  Giannakos, Amruth~N Kumar, Linda Ott, James Paterson, Michael~James Scott,
  Judy Sheard, et~al.
\newblock Introductory programming: a systematic literature review.
\newblock In {\em Proceedings companion of the 23rd annual ACM conference on
  innovation and technology in computer science education}, pages 55--106,
  2018.

\bibitem{marwan2020adaptive}
Samiha Marwan, Ge~Gao, Susan Fisk, Thomas~W Price, and Tiffany Barnes.
\newblock Adaptive immediate feedback can improve novice programming engagement
  and intention to persist in computer science.
\newblock In {\em Proceedings of the 2020 ACM conference on international
  computing education research}, pages 194--203, 2020.

\bibitem{mcguffin2020categories}
Michael~J McGuffin and Christopher~P Fuhrman.
\newblock Categories and completeness of visual programming and direct
  manipulation.
\newblock In {\em Proceedings of the 2020 International Conference on Advanced
  Visual Interfaces}, pages 1--8, 2020.

\bibitem{meyer1992eiffel}
Bertrand Meyer.
\newblock The eiffel programming language.
\newblock {\em See http://www. eiffel. com}, 1992.

\bibitem{naps2002exploring}
Thomas~L. Naps, Guido R\"{o}\ss{}ling, Vicki Almstrum, Wanda Dann, Rudolf
  Fleischer, Chris Hundhausen, Ari Korhonen, Lauri Malmi, Myles McNally, Susan
  Rodger, and J.~\'{A}ngel Vel\'{a}zquez-Iturbide.
\newblock Exploring the role of visualization and engagement in computer
  science education.
\newblock In {\em Working Group Reports from ITiCSE on Innovation and
  Technology in Computer Science Education}, ITiCSE-WGR '02, page 131–152,
  New York, NY, USA, 2002. Association for Computing Machinery.

\bibitem{resnick2009scratch}
Mitchel Resnick, John Maloney, Andr{\'e}s Monroy-Hern{\'a}ndez, Natalie Rusk,
  Evelyn Eastmond, Karen Brennan, Amon Millner, Eric Rosenbaum, Jay Silver,
  Brian Silverman, et~al.
\newblock Scratch: programming for all.
\newblock {\em Communications of the ACM}, 52(11):60--67, 2009.

\bibitem{scott2014direct}
Jeremy Scott, Philip~J Guo, and Randall Davis.
\newblock A direct manipulation language for explaining algorithms.
\newblock In {\em 2014 IEEE Symposium on Visual Languages and Human-Centric
  Computing (VL/HCC)}, pages 45--48. IEEE, 2014.

\bibitem{shneiderman2010designing}
Ben Shneiderman and Catherine Plaisant.
\newblock {\em Designing the user interface: strategies for effective
  human-computer interaction}.
\newblock Pearson Education India, 2010.

\bibitem{smith1975pygmalion}
David~Canfield Smith.
\newblock {\em Pygmalion: a creative programming environment.}
\newblock Stanford University, 1975.

\bibitem{stallman1993gnu}
Richard Stallman and Howard Gayle.
\newblock {\em GNU Emacs manual}, volume 675.
\newblock Free Software Foundation Cambridge, MA, 1993.

\bibitem{szydlowska2022python}
Justyna Szyd{\l}owska, Filip Miernik, Marzena~Sylwia Ignasiak, and Jakub
  Swacha.
\newblock Python programming topics that pose a challenge for students.
\newblock In {\em Third International Computer Programming Education Conference
  (ICPEC 2022)}, pages 7--1. Schloss Dagstuhl--Leibniz-Zentrum f{\"u}r
  Informatik, 2022.

\bibitem{thorgeirsson2024comparing}
Sverrir Thorgeirsson, Theo~B Weidmann, Karl-Heinz Weidmann, and Zhendong Su.
\newblock Comparing cognitive load among undergraduate students programming in
  python and the visual language algot.
\newblock In {\em Proceedings of the 55th ACM Technical Symposium on Computer
  Science Education V. 1}, pages 1328--1334, 2024.

\bibitem{trower2015blockly}
Jake Trower and Jeff Gray.
\newblock Blockly language creation and applications: Visual programming for
  media computation and bluetooth robotics control.
\newblock In {\em Proceedings of the 46th ACM Technical Symposium on Computer
  Science Education}, pages 5--5, 2015.

\bibitem{victor2012learnable}
Bret Victor.
\newblock Learnable programming.
\newblock {\em Worrydream.com}, 2012.

\bibitem{weidmann2022bridging}
Theo~B Weidmann, Sverrir Thorgeirsson, and Zhendong Su.
\newblock Bridging the syntax-semantics gap of programming.
\newblock In {\em Proceedings of the 2022 ACM SIGPLAN International Symposium
  on New Ideas, New Paradigms, and Reflections on Programming and Software},
  pages 80--94, 2022.

\bibitem{zeevaarders2021exploring}
Ad~Zeevaarders and Efthimia Aivaloglou.
\newblock Exploring the programming concepts practiced by scratch users: An
  analysis of project repositories.
\newblock In {\em 2021 IEEE Global Engineering Education Conference (EDUCON)},
  pages 1287--1295. IEEE, 2021.

\end{thebibliography}
\end{document}